\documentclass[]{trbunofficial}
\usepackage{graphicx}
\usepackage[usenames,dvipsnames]{xcolor}
\usepackage{amsmath} %
\usepackage{amssymb}  %
\usepackage{pstricks,pst-plot,psfrag}
\usepackage{booktabs}
\usepackage{subcaption}
\usepackage{amsthm} 
\usepackage{tikz}
\usepackage{bbold}
\usepackage[acronym]{glossaries}
\usepackage{comment}
\usepackage{lettrine}
\usepackage{import}
\usepackage{units}
\usepackage{mathtools}
\usepackage{dsfont}
\usepackage{booktabs}
\usepackage{siunitx}
\usepackage{multirow}

\definecolor{lightblue}{rgb}{0.60784,0.76078,0.90196}
\definecolor{darkblue}{rgb}{0.26667,0.44706,0.76863}
\definecolor{lightgreen}{rgb}{0.66275,0.81569,0.55686}
\definecolor{darkgreen}{rgb}{0.43922,0.67843,0.27843}
\definecolor{orange}{rgb}{0.92941,0.49020,0.19216}
\definecolor{yellow}{rgb}{1.00000,0.75294,0.00000}
\definecolor{grey}{rgb}{0.64706,0.64706,0.64706}
\definecolor{purple}{rgb}{0.51373,0.23529,0.04706}
\newacronym{abk:amod}{AMoD}{Autonomous Mobility-on-Demand}
\newacronym{abk:iamod}{I-AMoD}{intermodal \gls{abk:amod}}
\newacronym{abk:av}{AV}{Autonomous Vehicle}
\newacronym{abk:camod}{C-\gls{abk:amod}}{Co-Design for Autonomous Mobility-on-Demand Systems}
\newacronym{abk:cdp}{CDP}{Co-Design Problem}
\newacronym{abk:dp}{DP}{Design Problem}
\newacronym{abk:ffcs}{FFCS}{free floating car sharing systems}
\newacronym{abk:ghg}{GHG}{greenhouse gas}
\newacronym{abk:mcfp}{MCFP}{multi-commodity flow problem}
\newacronym{abk:mcdp}{MCDP}{Monotone Co-Design Problem}
\newacronym{abk:mpc}{MPC}{model predictive control}
\newacronym{abk:spp}{SPP}{shortest path problem}
\newacronym{abk:kdspp}{k-dSPP}{k-disjoint \gls{abk:spp}}

\newcommand{\setOfArcsRoad}{\mathcal{A}_{\mathrm{R}}}

\newcommand{\arc}{(i,j)}

\usetikzlibrary{positioning,intersections, hobby, patterns, calc, decorations.pathmorphing, decorations.markings, shadows,shapes}
\tikzstyle{block} = [draw, rectangle, minimum height=2em, minimum width=3em]
\tikzstyle{block1} = [draw, rectangle, minimum height=1.5em, minimum width=2.5em]
\tikzstyle{blockDyn} = [draw, rectangle, minimum height=2.5em, minimum width=3.5em, align=center, inner sep=10pt, thick, fill=white, copy shadow={draw=black,fill=black,opacity=1,shadow xshift=0.5ex,shadow yshift=-0.5ex}]
\tikzstyle{blockAlg} = [draw, rectangle, minimum height=1.5em, minimum width=2.5em, align=center, inner sep=10pt, thick]
\tikzstyle{sum} = [draw,circle]
\tikzstyle{nodePre} = [circle, draw,inner sep=1pt,node contents={$\preceq$}]
\tikzstyle{nodePos} = [circle, draw,inner sep=1pt,node contents={$\posceq$}]
\tikzstyle{nodeProd} = [rectangle, draw,inner sep=4pt,node contents={$\times$}]
\tikzstyle{nodeSum} = [rectangle, draw,inner sep=4pt,node contents={$+$}]

\definecolor{DPgreen}{RGB}{34,139,34}

\DeclareSIUnit{\mph}{mph}

\newif\ifmargincomments %
\margincommentstrue

\ifmargincomments

\else

\fi

\AuthorHeaders{Zardini, Lanzetti, Salazar, Censi, Frazzoli, and Pavone}
\title{T\lowercase{owards a }C\lowercase{o-}D\lowercase{esign }F\lowercase{ramework for }F\lowercase{uture }M\lowercase{obility }S\lowercase{ystems}}
\author{%
	\textbf{Gioele Zardini} (Corresponding Author)\\
	Stanford University, Aeronautics and Astronautics\\
	496 Lomita Mall, Stanford, CA 94305-4035, USA\\
	E-mail: gzardini@ethz.ch\\
	\hfill\break%
      \textbf{Nicolas Lanzetti} \\
	Stanford University, Aeronautics and Astronautics\\
	496 Lomita Mall, Stanford, CA 94305-4035, USA\\
	E-mail: lnicolas@ethz.ch\\
    \hfill\break%
    \textbf{Mauro Salazar, Ph.D.}\\
    Stanford University, Aeronautics and Astronautics\\
    496 Lomita Mall, Stanford, CA 94305-4035, USA\\
    E-mail: samauro@stanford.edu\\
    \hfill\break%
    \textbf{Andrea Censi, Ph.D.}\\
    ETH Zurich, Institute for Dynamic Systems and Control\\
    Sonneggstrasse 3, 8092 Zurich, ZH-Switzerland\\
    E-mail: acensi@ethz.ch\\
    \hfill\break%
    \textbf{Emilio Frazzoli, Ph.D.}\\
    ETH Zurich, Institute for Dynamic Systems and Control\\
    Sonneggstrasse 3, 8092 Zurich, ZH-Switzerland\\
    E-mail: emilio.frazzoli@idsc.mavt.ethz.ch\\
    \hfill\break%
    \textbf{Marco Pavone, Ph.D.}\\
    Stanford University, Aeronautics and Astronautics\\
    496 Lomita Mall, Stanford, CA 94305-4035, USA\\
    E-mail: pavone@stanford.edu
}

\begin{document}
\maketitle 
\section{Abstract}

The design of Autonomous Vehicles (AVs) and the design of AVs-enabled mobility systems are closely coupled. Indeed, knowledge about the intended service of AVs would impact their design and deployment process, whilst insights about their technological development could significantly affect transportation management decisions.
This calls for tools to study such a coupling and co-design AVs and AVs-enabled mobility systems in terms of different objectives.
In this paper, we instantiate a framework to address such co-design problems.
In particular, we leverage the recently developed theory of co-design to frame and solve the problem of designing and deploying an intermodal Autonomous Mobility-on-Demand system, whereby AVs service travel demands jointly with public transit, in terms of fleet sizing, vehicle autonomy, and public transit service frequency. Our framework is modular and compositional, allowing to describe the design problem as the interconnection of its individual components and to tackle it from a system-level perspective.
Moreover, it only requires very general monotonicity assumptions and it naturally handles multiple objectives, delivering the \emph{rational} solutions on the Pareto front and thus enabling policy makers to select a solution through ``political'' criteria.
To showcase our methodology, we present a real-world case study for Washington D.C., USA.
Our work suggests that it is possible to create user-friendly optimization tools to systematically assess the costs and benefits of interventions, and that such analytical techniques might gain a momentous role in policy-making in the future.

\hfill\break%
\noindent\textit{Keywords}: Co-Design, Autonomous Vehicles, Future Mobility Systems.

\newpage

\section{Introduction}

Arguably, the current design process for \glspl{abk:av} largely suffers from the lack of clear, specific requirements in terms of the service such vehicles will be providing.  Yet, knowledge about their intended service (e.g., last-mile versus point-to-point travel) might dramatically impact how the AVs are designed, and, critically, significantly ease their development process. For example, if for a given city we knew that for an effective on-demand mobility system autonomous cars only need to drive up to \unit[25]{mph} and only on relatively easy roads, their design would be greatly simplified and their deployment could certainly be accelerated.
At the same time, from the system-level perspective of transportation management, knowledge about the trajectory of technology development for  \glspl{abk:av} would certainly impact decisions on infrastructure investments and provision of service. 
In other words, the design of the \glspl{abk:av} and the design of a mobility system leveraging \glspl{abk:av} are intimately {\em coupled}. 
This calls for methods  to reason about such a coupling, and in particular to \emph{co-design} the \glspl{abk:av} and the associated \glspl{abk:av}-enabled mobility system. A key requirement in this context is to be able to account  for a range of heterogeneous objectives that are often not directly comparable (consider, for instance, travel time and emissions). 

Accordingly, the goal of this paper is to lay the foundations for a framework through which one can co-design   future \glspl{abk:av}-enabled mobility systems. 
Specifically, we show how one can leverage the recently developed mathematical theory of co-design~\cite{Censi2015,Censi2016,Censi2017b}, which provides a general methodology to co-design complex systems in a modular and compositional fashion.
This tool delivers the set of rational design solutions lying on the Pareto front, allowing to reason about the costs and benefits of the individual design options.
The framework is instantiated in the setting of co-designing intermodal  \gls{abk:amod} systems~\cite{SalazarLanzettiEtAl2019}, whereby fleets of self-driving vehicles provide on-demand mobility jointly with public transit. Aspects that are subject to co-design include fleet size, vehicle-specific characteristics for the \glspl{abk:av}, and public transit service frequency.

\emph{Literature Review:} The design of mobility systems can be divided in classic urban transportation network design problems and more recent design problems for \gls{abk:amod} systems.
The first research stream was reviewed in~\cite{FarahaniMiandoabchiEtAl2013} and can be divided in road~\cite{WangLo2010,CongDeSchutterEtAl2015}, public transit~\cite{GuihaireHao2008,CiprianiGoriEtAl2012}, and multi-modal~\cite{MiandoabchiFarahaniEtAl2012} network design problems which can be classified as \emph{strategic} long-term infrastructure modification decisions such as building new streets, \emph{tactical} infrastructure-allocation problems on lanes allocation and public transit service frequency, and \emph{operational} short-term scheduling problems.
They are usually formulated as bilevel problems whereby the upper-level is related to the policy in discussion and the lower-level is concerned with solving the trip assignment problem by computing the user equilibrium or the system optimum under given congestion and demand models.
Overall, these problems are solved with non-convex and combinatorial mathematical methods, heuristics, and metaheuristics consisting of gradient-free optimization algorithms. 
Often, the problems are formulated with a unique objective or, in the case of multi-objective settings, the different objectives are reduced to a unique one through monetary metrics precluding Pareto solutions.
The design of \gls{abk:amod} systems mostly considers their fleet sizing.
In particular, it was studied through simulations in~\cite{BarriosGodier2014, FagnantKockelman2018} and with analytical methods in~\cite{SpieserTreleavenEtAl2014}, whilst in~\cite{ZhangSheppardEtAl2018} the problem was combined with the charging infrastructure sizing and placement problem and solved using mixed integer linear programming techniques. 
The fleet sizing and vehicle allocation problem for conventional vehicles was presented in~\cite{BeaujonTurnquist1991}.
The fleet size and pricing scheme of Mobility-on-Demand systems was designed in~\cite{LiuBansalEtAl2018}  with Bayesian optimization, whereas in~\cite{SayarshadTavakkoli-Moghaddam2010} a rail-car fleet sizing problem was solved with simulated annealing.
More recently, the joint design of multimodal transit networks and \gls{abk:amod} systems was formulated in~\cite{PintoHylandEtAl2019} as a bilevel optimization problem and solved with heuristic methods.
Overall, to the best of the authors' knowledge, most design methods for \gls{abk:amod} rely either on simulation-based approaches or nonlinear and combinatorial optimization techniques, and do not study \glspl{abk:av}-specific characteristics such as the achievable vehicle speed.
In conclusion, the frameworks proposed for the design of mobility systems mainly have a fixed problem-specific structure and are thus non-modular.
Moreover, they do not deliver a Pareto front of solutions, focusing on a unique objective.

\emph{Statement of Contribution:}
In this paper we lay the foundations for the systematic study of the design of \glspl{abk:av}-enabled mobility systems. Specifically, we leverage the mathematical theory of co-design~\cite{Censi2015} to devise a framework to study the design of \gls{abk:iamod} systems in terms of fleet characteristics and public transit service, enabling the computation of the \emph{rational} solutions lying on the Pareto front of minimal travel time, transportation costs, and emissions.
Our framework allows to structure the design problem in a modular way, in which each different transportation option can be ``plugged in'' in a larger model. 
Each model has minimal assumptions: Rather than properties such as linearity and convexity, we ask for very general monotonicity assumptions. For example, we assume that the cost of automation increases monotonically with the speed achievable by the self-driving car.
We are able to obtain the full Pareto front of \emph{rational} solutions, or, given more ``political'' criteria, to weigh incomparable costs (such as travel time and emissions), to present one optimal solution to stakeholders, such as \glspl{abk:av} companies and municipalities.
We consider the real-world case study for Washington~D.C., to showcase our methodology. We show how, given the model, we can easily formulate and answer several questions regarding the introduction of new technologies and investigate possible infrastructure interventions. A proceedings version of this work has been presented at the 23rd IEEE Intelligent Transportation Systems Conference~\cite{Zardini2020}.

\subsection{Organization}
The remainder of this paper is structured as follows: Section~\ref{sec:background} presents the mathematical background of co-design.
Section~\ref{sec:model} presents the co-design problem for \glspl{abk:av}-enabled mobility systems.
We showcase our approach with real-world case studies for Washington D.C., USA, in Section~\ref{sec:results}. Section~\ref{sec:conclusion} concludes the paper with a discussion and an overview on future research directions.

\section{Mathematical Background}\label{sec:background}
In this section, we present the basics of partial order theory and the mathematical theory of co-design. The interested reader is referred to~\cite{Censi2015,Censi2016,Censi2017b}.

\vspace{0.4 \baselineskip}
\noindent \textit{Partial Order Theory} 
\vspace{0.1 \baselineskip}

\noindent Consider a set $\mathcal{P}$ and a partial order $\preceq_\mathcal{P}$, defined as a reflexive, antisymmetric, and transitive relation~\cite{DaveyPriestley2002}. Then, $\mathcal{P}$ and $\preceq_\mathcal{P}$ define the partially ordered set (poset) $\langle \mathcal{P}, \preceq_\mathcal{P} \rangle$. 
The least and maximum elements of a poset are called bottom and top, and are denoted by $\bot_\mathcal{P}$ and $\top_\mathcal{P}$, respectively. A set $S\subseteq \mathcal{P}$ is \emph{directed} if each pair of elements $x,y \in S$ has an upper bound. A poset is a \emph{directed complete partial order} (DCPO) if each of its directed subsets has a top, and it is a \emph{complete partial order} (CPO) if it has a bottom as well.  A \emph{chain} is a subset $S\subseteq \mathcal{P}$ where all elements are comparable, i.e., for $x,y\in S, \ x\preceq_\mathcal{P} y$ or $y\preceq_\mathcal{P} x$. Conversely, an \emph{antichain} is a subset $S\subseteq \mathcal{P}$ where no elements are comparable, i.e., for $x,y \in S, \ x \preceq_\mathcal{P} y$ implies $x=y$. A map $g:\mathcal{P}\rightarrow \mathcal{Q}$ between two posets is \emph{monotone} iff $x\preceq_\mathcal{P} y$ implies $g(x) \preceq_\mathcal{Q} g(y)$.

\vspace{0.4 \baselineskip}
\noindent \textit{Mathematical Theory of Co-Design} 
\vspace{0.1 \baselineskip}

\noindent As in \cite{Censi2015}, we abstract a \gls{abk:dp} as a monotone map $h$ between provided \emph{functionalities} and the antichain of required \emph{resources}, represented as elements of the CPOs $\langle \mathcal{F},\preceq_{\mathcal{F}}\rangle$, $\langle \mathcal{R},\preceq_{\mathcal{R}}\rangle$, respectively. 
Different to classical approaches, mostly relying on properties such as continuity, linearity, or convexity, our approach only requires monotone relations between the antichain of resources and the functionalities.
The \gls{abk:cdp} of the full system results then from the interconnection, typically given in form of a graph, of the \gls{abk:dp} of its individual components. Indeed, this allows to describe complex models in a modular and compositional fashion.
We focus on the problem of finding the antichain of all \emph{rational} resources $\mathsf{r}_1,\ldots,\mathsf{r}_N\in\mathcal{R}$ which provide a given functionality $\mathsf{f}\in \mathcal{F}$. Nevertheless, the framework can readily accommodate alternative problem formulations, such as finding the antichain of all rational functionalities $\mathsf{f}_1,\hdots,\mathsf{f}_N \in \mathcal{F}$ which are provided given a resource $\mathsf{r} \in \mathcal{R}$. Rather than computing a single solution, this method provides therefore an antichain, or equivalently, a set of incomparable rational decisions. 

\section{Co-Design of AV-enabled Mobility Systems}\label{sec:model}
In this section, we instantiate our proposed co-design framework in the setting whereby a central, social welfare maximizing authority (e.g., a central authority) strives to co-design a mobility system comprising AMoD and public transportation, in terms of \gls{abk:av} fleet size, vehicle-specific characteristics, and public transit service frequency. This rather idealized setting serves a number of purposes: First, it grounds our co-design framework within a concrete urban transportation design problem. Second, the insights we derive can be interpreted as upper bounds on the performance gains one might achieve via co-design. Third, this setting subsumes as special cases the co-design of \gls{abk:av} and AMoD services (of interest, e.g., to \gls{abk:av} and mobility-as-a-service companies alike) and the co-design of AMoD and intermodal transportation systems (of interest, e.g., to municipal authorities). Fourth, it provides a starting point to address the more challenging (and more realistic) setting whereby multiple stakeholders, with different objectives, might converge via co-design to an optimized transportation system. 

We start by describing in Section 4.1 the urban transportation design problem we want to address in this paper; we then present its associated co-design framework in Section 4.2.
\subsection{Intermodal \gls{abk:amod} Framework}
\label{subsec:iamod}
In this section, we present the \gls{abk:iamod} framework from~\cite{SalazarLanzettiEtAl2019} used to describe our setting.
We adopt a mesoscopic planning perspective and formulate the problem as a multi-commodity network flow problem, whereby we allow customers to be routed in an intermodal fashion.
\subsubsection{Multi-Commodity Flow Model}
\begin{figure}[hbt]
	\begin{center}
		\begin{subfigure}[b]{0.49\textwidth}
			\centering
			\includegraphics[width= \linewidth]{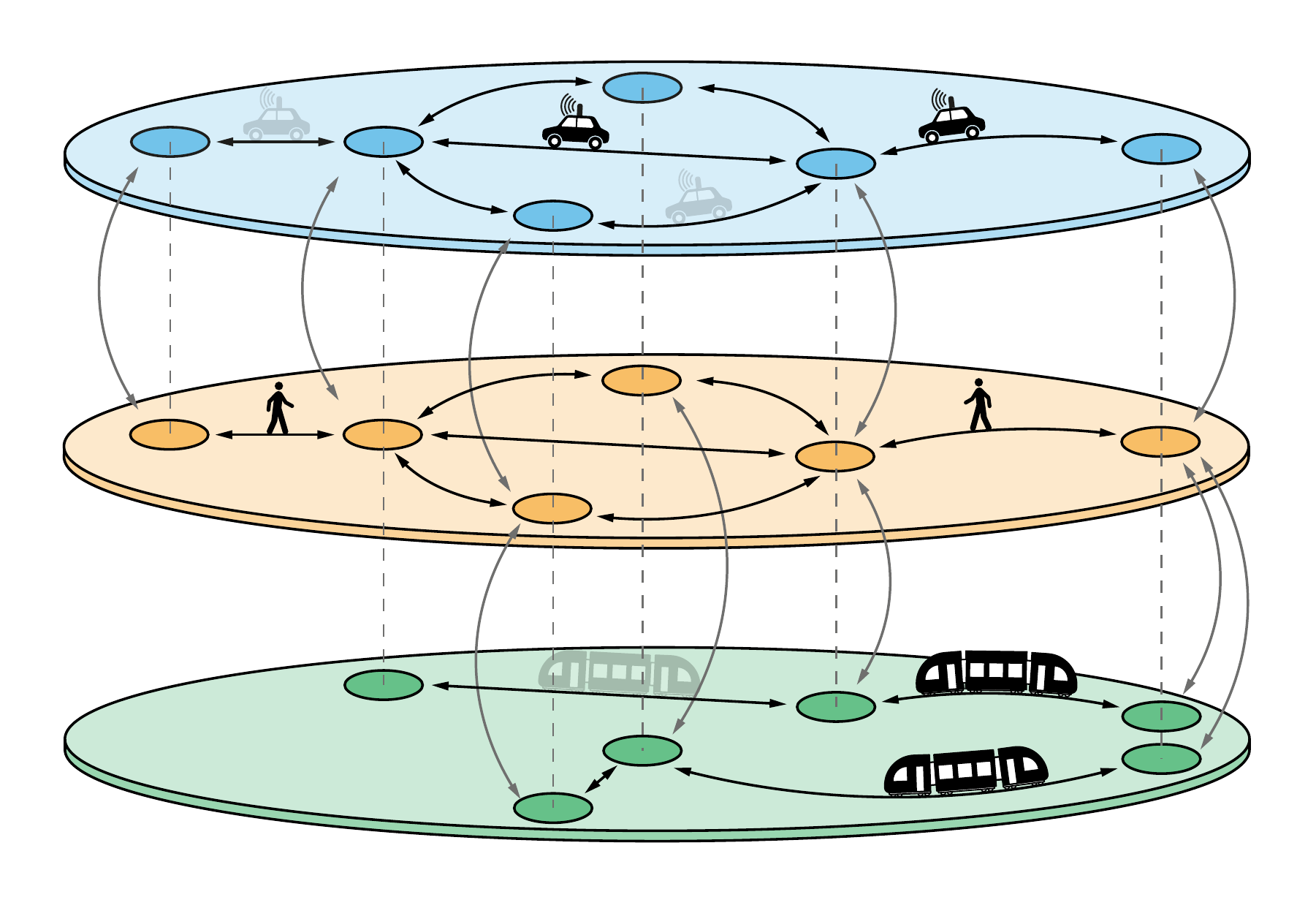}
			\caption{Intermodal \gls{abk:amod} network.}
			\label{fig:digraph}
		\end{subfigure}
		\begin{subfigure}[b]{0.49\textwidth}
			\centering
			\includegraphics[width= \linewidth]{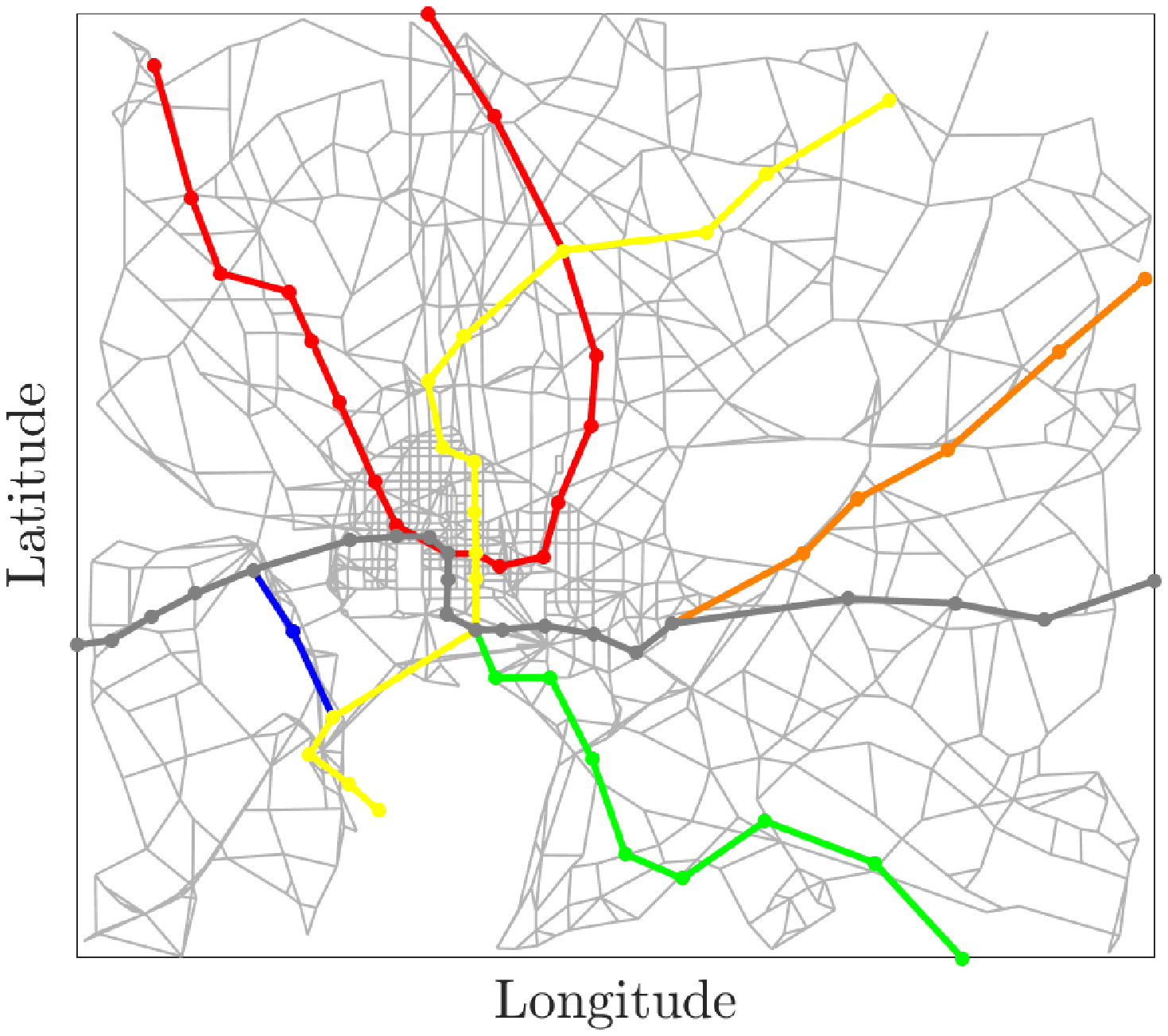}
			\caption{Washington D.C. road and subway graphs.}
			\label{fig:dc}
		\end{subfigure}
		\caption{(a) The intermodal AMoD network consists of a road, a walking, and a public transportation digraph. The coloured circles represent stops or intersections and the black arrows denote road links, pedestrian pathways, or public transit arcs. The dotted lines represent nodes which are close geographically, while the grey arrows represent the mode-switching arcs connecting them. (b) The Washington D.C. transportation network, consisting of the MetroRail subway lines and the city roads.}
		\label{fig:iamod}
	\end{center}
\end{figure}

The transportation system and its different modes are modeled using the digraph $\mathcal{G}=\left( \mathcal{V}, \mathcal{A} \right)$, shown in Figure \ref{fig:digraph}.
It is composed of a set of nodes $\mathcal{V}$ and a set of arcs $\mathcal{A}\subseteq \mathcal{V}\times \mathcal{V}$. Specifically, it contains a road network layer $\mathcal{G}_\mathrm{R}=\left( \mathcal{V}_\mathrm{R}, \mathcal{A}_\mathrm{R}\right)$, a public transportation layer $\mathcal{G}_\mathrm{P}=\left( \mathcal{V}_\mathrm{P}, \mathcal{A}_\mathrm{P}\right)$, and a walking layer $\mathcal{G}_\mathrm{W}=\left( \mathcal{V}_\mathrm{W}, \mathcal{A}_\mathrm{W}\right)$. The road network is characterized through intersections $i \in \mathcal{V}_\mathrm{R}$ and road segments $(i,j)\in \mathcal{A}_\mathrm{R}$. Similarly, public transportation lines are modeled through station nodes $i\in \mathcal{V}_\mathrm{P}$ and line segments $(i,j)\in \mathcal{A}_\mathrm{P}$. The walking network contains walkable streets $(i,j) \in \mathcal{A}_\mathrm{W}$ connecting intersections $i\in \mathcal{V}_\mathrm{W}$. Our model allows mode-switching arcs $\mathcal{A}_\mathrm{C} \subseteq \mathcal{V}_\mathrm{R} \times \mathcal{V}_\mathrm{W}\cup \mathcal{V}_\mathrm{W} \times \mathcal{V}_\mathrm{R}\cup \mathcal{V}_\mathrm{P}\times \mathcal{V}_\mathrm{W} \cup \mathcal{V}_\mathrm{W}\times \mathcal{V}_\mathrm{P}$ connecting the road and the public transportation layers to the pedestrian layer. Consequently, $\mathcal{V}=\mathcal{V}_\mathrm{W} \cup \mathcal{V}_\mathrm{R} \cup \mathcal{V}_\mathrm{P}$ and $\mathcal{A}=\mathcal{A}_\mathrm{W} \cup \mathcal{A}_\mathrm{R} \cup \mathcal{A}_\mathrm{P} \cup \mathcal{A}_\mathrm{C}$.
Consistently with the structural properties of road and walking networks in urban environments, we assume the graph $\mathcal{G}$ to be strongly connected.
We model a travel request $\rho$ as a triple $(o,d,\alpha) \in \mathcal{V}\times \mathcal{V}\times \mathbb{R}_+$, described by its origin node $o$, its destination node $d$, and its request rate $\alpha >0$, namely, how many customers want to travel from $o$ to $d$ per unit time.
We assume that the origin and destination vertices of the $M$ requests lie in the walking digraph, i.e.,  $o_m,d_m \in \mathcal{V}_\mathrm{W}$ for all $m \in \mathcal{M}\coloneqq\{1,\ldots,M\}$.

The flow $f_m(i,j)$ represents the number of customers per unit time traversing arc $(i,j)\in \mathcal{A}$ and satisfying a travel request $m$. Furthermore, $f_0(i,j)$ denotes the flow of empty vehicles on road arcs $(i,j)\in \mathcal{A}_\mathrm{R}$, accounting for rebalancing flows of \gls{abk:amod} vehicles between a customer's drop-off and the next customer's pick-up. Assuming the vehicles to carry one customer at a time, the flows satisfy
\begin{linenomath}
\begin{subequations}
\label{eq:flowconstotal}
\begin{align}
\label{eq:flowconsa}
&\sum_{i:(i,j)\in \mathcal{A}} f_m(i,j) + \mathbb{1}_{j=o_m}\cdot \alpha_m = \sum_{k:(j,k)\in \mathcal{A}} f_m(j,k)+\mathbb{1}_{j=d_m} \cdot \alpha_m,\quad \forall m \in \mathcal{M}, j\in \mathcal{V}
\\
\label{eq:flowconsb}
&\sum_{i:(i,j)\in \mathcal{A}_\mathrm{R}} \left( f_0(i,j)+\sum_{m\in \mathcal{M}}f_m(i,j)\right)= \sum_{k:(j,k)\in \mathcal{A}_\mathrm{R}}\left( f_0(j,k)+\sum_{m\in \mathcal{M}} f_m(j,k) \right),\quad  \forall j \in \mathcal{V}_\mathrm{R}
\\
\label{eq:nonnega}
&f_m(i,j)\geq 0, \quad \forall m \in \mathcal{M}, (i,j)\in \mathcal{A}
\\
\label{eq:nonnegb}
&f_0(i,j)\geq 0, \quad \forall (i,j)\in \mathcal{A}_\mathrm{R},
\end{align}
\end{subequations}
\end{linenomath}
where $\mathbb{1}_{j=x}$ denotes the boolean indicator function. Specifically, \eqref{eq:flowconsa} guarantees flows conservation for every transportation demand, and \eqref{eq:flowconsb} preserves flow conservation for vehicles on every road node. Combining conservation of customers~\eqref{eq:flowconsa} with the conservation of vehicles~\eqref{eq:flowconsb} guarantees rebalancing vehicles to match the demand.
Finally, \eqref{eq:nonnega}, \eqref{eq:nonnegb} ensure non-negativity of flows.
\subsubsection{Travel Time and Travel Speed}
\label{sec:speedcong}
The variable $t_{ij}$ denotes the time needed to traverse an arc $(i,j)$ of length $s_{ij}$. We assume a constant walking speed on pedestrian arcs and infer travel times on public transportation arcs from the public transit schedules. Considering that the public transportation system at node $j$ operates with the frequency $\varphi_{j}$, switching from a pedestrian vertex $i$ to a public transit station $j$ takes, on average, $t_\mathrm{WS}+1/2\varphi_j$, where $t_\mathrm{WS}$ is a constant sidewalk-to-station travel time. 
We assume that the average waiting time for \gls{abk:amod} vehicles is $t_\mathrm{WR}$ and that switching from the road graph and the public transit graph to the pedestrian graph takes the transfer times $t_\mathrm{RW}$ and $t_\mathrm{SW}$, respectively.
While each road arc $(i,j) \in \mathcal{A}_\mathrm{R}$ is characterized by a speed limit $v_{\mathrm{L},ij}$, \glspl{abk:av} safety protocols impose a maximum achievable velocity $v_{\mathrm{a}}$. 
In order to prevent too slow and therefore dangerous driving behaviours ~\cite{Dahl2018}, we only consider road arcs through which the \glspl{abk:av} can drive at least at a fraction $\beta$ of the speed limit: Arc $(i,j)\in\mathcal{A}_\mathrm{R}$ is kept in the road network if and only if 
\begin{linenomath}
\begin{equation}
    \label{eq:droparcs}
    v_{\mathrm{a}} \geq \beta \cdot v_{\mathrm{L},ij},
\end{equation}
\end{linenomath}
where $\beta\in (0,1]$.
We set the velocity of all arcs fulfilling condition \eqref{eq:droparcs} to $v_{ij}=\min\{v_{\mathrm{a}},v_{\mathrm{L},ij}\}$ and compute the travel time to traverse them as
\begin{linenomath}
\begin{equation}
    t_{ij} = \frac{s_{ij}}{v_{ij}}.
\end{equation}
\end{linenomath}

\subsubsection{Road Congestion}

In our setting, we assume that each road arc $(i,j)\in\mathcal{A}_\mathrm{R}$ is subject to a baseline usage $u_{ij}$, capturing the presence of exogenous traffic (e.g., private vehicles), and that it has a nominal capacity $c_{ij}$.
Furthermore, we assume that the central authority operates the \gls{abk:amod} fleet such that vehicles travel at free-flow speed throughout the road network of the city, meaning that the total flow on each road link must be below the link's capacity. Therefore, we capture congestion effects with the threshold model
\begin{linenomath}
\begin{equation}
\label{eq:capacity}
f_0(i,j)+\sum_{m\in \mathcal{M}} f_m(i,j)+u_{ij} \leq c_{ij} \quad \forall (i,j) \in \mathcal{A}_\mathrm{R}.
\end{equation}
\end{linenomath}

\subsubsection{Energy Consumption}
We compute the energy consumption of the \glspl{abk:av} for each road link considering an urban driving cycle, scaled so that the average speed $v_\mathrm{avg,cycle}$ matches the free-flow speed on the link $s_{ij}/t_{ij}$, and scale the energy consumption with the length of the arc $s_{ij}$ as
\begin{linenomath}
\begin{equation}
e_{ij} = e_\mathrm{cycle}\cdot\frac{s_{ij}}{s_\mathrm{cycle}}\quad\forall\arc \in \setOfArcsRoad.
\end{equation}
\end{linenomath}
For the public transportation system, we assume a constant energy consumption per unit time. This approximation is acceptable in urban environments, as the operation of the public transportation system is independent from the number of customers serviced, and its energy consumption is therefore customers-invariant.

\subsubsection{Fleet Size}
We consider a fleet of $n_\mathrm{v,max}$ \glspl{abk:av}. In a time-invariant setting, the number of vehicles on arc $(i,j)$ is expressed as the multiplication of the total vehicles flow on the arc and its travel time. Therefore, we constrain the number of vehicles employed as
\begin{linenomath}
\begin{equation}
    \label{eq:fleetsize}
    n_\mathrm{v,e} = \sum_{(i,j)\in \mathcal{A}_\mathrm{R}} \left( f_0(i,j)+\sum_{m\in\mathcal{M}}f_m(i,j)\right)\cdot t_{ij}
    \leq n_\mathrm{v,max}.
\end{equation}
\end{linenomath}

\subsubsection{Discussion}
\label{subsubsec:discussion iamod}
A few comments are in order.
First, we assume the demand to be time-invariant and allow flows to have fractional values. This assumption is in line with the \emph{mesoscopic} and system-level \emph{planning} perspective of our study.
Second, we model congestion using a threshold model. This approach can be interpreted as the municipal authority not allowing the \gls{abk:amod} vehicles to exceed the critical density of the flows on road arcs, so that cars can be assumed to travel at free flow speed~\cite{Daganzo2008}. This way, we can assume that the route planning of \gls{abk:amod} vehicles does not influence the exogenous traffic.
Finally, in line with the status quo, we allow \gls{abk:amod} vehicles to transport one customer at the time ~\cite{PIM2012}.
For further discussions on our modeling assumptions, we refer the readers to~\cite{SalazarRossiEtAl2018,SalazarLanzettiEtAl2019}.

\subsection{Co-Design Framework}
We integrate the \gls{abk:iamod} framework presented in Section~\ref{subsec:iamod} in the co-design formalism, allowing the decoupling of the \gls{abk:cdp} of a complex system in the \gls{abk:dp} of its individual components in a modular, compositional, and systematic fashion. 
We aim to compute the antichain of resources, quantified in terms of costs, average travel time per trip, and emissions required to provide the mobility service to a set of customers.
In order to achieve this, we decouple the \gls{abk:cdp} in the \glspl{abk:dp} of the individual \gls{abk:av} (Section \ref{sec:vehdp}) and of the \glspl{abk:av} fleet (Section \ref{sec:iamodp}) as well as of the public transportation system (Section \ref{sec:subdp}).
Their interconnection is presented in Section \ref{sec:mcdp}.

\subsubsection{The Autonomous Vehicle Design Problem}
\label{sec:vehdp}
The \gls{abk:av} \gls{abk:dp} (Figure \ref{fig:vehdp}) consists of selecting the maximal speed of the \glspl{abk:av}. 
Under the rationale that driving safely at higher speed requires more advanced sensing and algorithmic capabilities, we model the achievable speed of the \glspl{abk:av} $v_\mathrm{a}$ as a monotone function of the vehicle fixed costs $C_\mathrm{v,f}$ (resulting from the cost of the vehicle $C_\mathrm{v,v}$ and the cost of its automation $C_\mathrm{v,a}$) and the mileage-dependent operational costs $C_\mathrm{v,o}$ (accounting for maintenance, cleaning, energy consumption, depreciation, and opportunity costs ~\cite{Mas-ColellWhinstonEtAl1995}).
In this setting, the \gls{abk:av} DP provides the functionality $v_\mathrm{a}$ and requires the resources $C_\mathrm{v,f}$ and $C_\mathrm{v,o}$. Consequently, the functionality space is $\mathcal{F}_\mathrm{v}=\overline{\mathbb{R}}_+$ (in \unit[]{mph}), and the resources space is $\mathcal{R}_\mathrm{v}=\overline{\mathbb{R}}_+\times \overline{\mathbb{R}}_+$ (in $\unit[]{USD} \times \unitfrac[]{USD}{mile}$). 

\subsubsection{The Subway Design Problem}
\label{sec:subdp}
We design the public transit infrastructure (Figure \ref{fig:subdp}) by means of the service frequency introduced in Section ~\ref{sec:speedcong}. 
Specifically, we assume the service frequency $\varphi_{j}$ to scale linearly with the size of the train fleet $n_\mathrm{s}$ as
\begin{linenomath}
\begin{equation}
\frac{\varphi_{j}}{\varphi_{j,\mathrm{baseline}}}=\frac{n_\mathrm{s}}{n_\mathrm{s,baseline}}.
\end{equation}
\end{linenomath}
We relate a train fleet of size $n_\mathrm{s}$ to the fixed costs $C_\mathrm{s,f}$ (accounting for train and infrastructural costs) and to the operational costs $C_\mathrm{s,o}$ (accounting for energy consumption, vehicles depreciation, and train operators' wages).
Given the passengers-independent public transit operation in today's cities, we reasonably assume the operational costs $C_\mathrm{s,o}$ to be mileage independent and to only vary with the size of the fleet.
Formally, the number of acquired trains $n_\mathrm{s,a}=n_\mathrm{s}-n_\mathrm{s,baseline}$ is a functionality, whereas $C_\mathrm{s,f}$ and $C_\mathrm{s,o}$ are resources. The functionality space is $\mathcal{F}_\mathrm{s}=\overline{\mathbb{N}}$ and the resources space is $\mathcal{R}_\mathrm{s}=\overline{\mathbb{R}}_+ \times \overline{\mathbb{R}}_+$ (in $\unit[]{USD} \times \unitfrac[]{USD}{year}$).

\subsubsection{The \gls{abk:iamod} Optimization Framework Design Problem}
\label{sec:iamodp}
\begin{figure}[!h]
    \begin{center}
    \begin{subfigure}[b]{0.32\textwidth}
    \centering
    \scalebox{0.7}{
    \begin{tikzpicture}[auto,node distance=2cm]
    \node [block, scale=1.5] (car){Vehicle};
    \node [circle,below=1cm of car] {};
    \draw [color=DPgreen,thick] ($(car.east)+(0,0)$) -- ($(car.east)+(1.5,0)$)node[pos=0.5,above]{$v_\mathrm{a}$}node[pos=0,right,circle,color=DPgreen,draw,fill=DPgreen,scale=0.4,solid] {};
    
    \draw [-, color=red,thick, dashed] ($(car.west)+(0,0.4)$) -- ($(car.west)+(-1.5,0.4)$)node[pos=0.5,above]{$C_\mathrm{v,f}$} node[pos=0,left,circle,color=red,draw,fill=red,scale=0.4,solid] {};
    \draw [-, color=red,thick, dashed] ($(car.west)+(0,-0.4)$) -- ($(car.west)+(-1.5,-0.4)$)node[pos=0.5,above]{$C_\mathrm{v,o}$}
    node[pos=0,left,circle,color=red,draw,fill=red,scale=0.4,solid] {};
    \end{tikzpicture}}
    \caption{Design problem of the autonomous vehicles.}
    \label{fig:vehdp}
    \end{subfigure}
    ~
    \begin{subfigure}[b]{0.32\textwidth}
    \scalebox{0.7}{
    \begin{tikzpicture}[auto,node distance=2cm]
    \node [block, scale=1.5] (car){Subway};
    \node [circle,below=1cm of car] {};
    \draw [color=DPgreen,thick] ($(car.east)+(0,0)$) -- ($(car.east)+(1.5,0)$)node[pos=0.5,above]{$n_\mathrm{s,a}$} node[pos=0,right,circle,color=DPgreen,draw,fill=DPgreen,scale=0.4,solid] {};
    \draw [-, color=red,thick,dashed] ($(car.west)+(0,0.4)$) -- ($(car.west)+(-1.5,0.4)$)node[pos=0.5,above]{$C_\mathrm{s,f}$} node[pos=0,left,circle,color=red,draw,fill=red,scale=0.4,solid] {};
    \draw [-, color=red,thick,dashed] ($(car.west)+(0,-0.4)$) -- ($(car.west)+(-1.5,-0.4)$)node[pos=0.5,above]{$C_\mathrm{s,o}$}node[pos=0,left,circle,color=red,draw,fill=red,scale=0.4,solid] {};
    \end{tikzpicture}}
    \centering
    \caption{Design problem of the subway infrastructure.}
    \label{fig:subdp}
    \end{subfigure}
    ~    
    \begin{subfigure}[b]{0.32\textwidth}
    \begin{center}
    \scalebox{0.7}{
    \begin{tikzpicture}[auto,node distance=2cm]
    \node [block, scale=1.5,minimum width = 4.5cm] (amod){I-AMoD};
    \draw[color=DPgreen,thick] ($(amod.north)+(0,0.75)$) -- ($(amod.north)+(0,0)$) node[pos=0.5,right]{$\alpha_\mathrm{tot}$}  node[pos=1,above,circle,color=DPgreen,draw,fill=DPgreen,scale=0.4] {};
    \draw[color=red,thick,dashed] ($(amod.south)+(-0.75,-0.75)$) -- ($(amod.south)+(-0.75,0)$) node[pos=0.5,left]{$v_\mathrm{a}$}  node[pos=1,below,circle,color=red,draw,fill=red,scale=0.4,solid] {};
    \draw[color=red, thick,dashed] ($(amod.south)+(0.75,-0.75)$) -- ($(amod.south)+(0.75,0)$) node[pos=0.5,left]{$n_\mathrm{v,max}$}  node[pos=1,below,circle,color=red,draw,fill=red,scale=0.4,solid] {};
    \draw[color=red,thick,dashed] ($(amod.south)+(-1.75,-0.75)$) -- ($(amod.south)+(-1.75,0)$) node[pos=0.5,left]{$n_\mathrm{s,a}$}  node[pos=1,below,circle,color=red,draw,fill=red,scale=0.4,solid] {};
    \draw[color=red,thick,dashed] ($(amod.south)+(1.75,-0.75)$) -- ($(amod.south)+(1.75,0)$) node[pos=0.5,left]{$t_\mathrm{avg}$}  node[pos=1,below,circle,color=red,draw,fill=red,scale=0.4,solid] {};
    \draw[color=red,thick,dashed] ($(amod.south)+(-2.75,-0.75)$) -- ($(amod.south)+(-2.75,0)$) node[pos=0.5,left]{$s_\mathrm{v,tot}$}  node[pos=1,below,circle,color=red,draw,fill=red,scale=0.4,solid] {};
    \draw[color=red,thick,dashed] ($(amod.south)+(2.75,-0.75)$) -- ($(amod.south)+(2.75,0)$) node[pos=0.5,right]{$m_\mathrm{CO_2,v,tot}$}  node[pos=1,below,circle,color=red,draw,fill=red,scale=0.4,solid] {};
    \end{tikzpicture}}
    \end{center}
    \caption{Design problem of the \gls{abk:iamod} system.}
    \label{fig:amoddp}
    \end{subfigure}
    \vspace{1cm}
    \begin{subfigure}[h]{\textwidth}
    \begin{center}
    
    \scalebox{0.7}{
    \begin{tikzpicture}[auto,node distance=0cm]    
    \node [block, scale=1.5,minimum width = 4.5cm,anchor=west] (amod)  {I-AMoD};
    \node [block,below left = 2.5 cm and 1.25cm of amod,  scale=1.5] (veh){Vehicle};
    \node [block,below right = 2.5 cm and 1.25cm of amod, scale=1.5] (sub){Subway};
    \node [nodePre, right = 1.6cm of veh,name=pre];
    \node [nodePre, left = 2.6cm of sub,name=pre2];
    \node [nodePre, right = 0.44cm of pre,name=pre3];
    \node [nodeProd, below = 1 cm of pre3, name=prod];
    \node [nodePre, left = 2.75 cm of prod, name=pre4];
    \node [nodePre, below left = 0.5 cm and 1.1 cm of pre4, name=pre5];
    \node [nodePre, right = 0.44 cm of pre3, name=pre6];
    \node [nodeProd, below = 1.9cm of pre6, name=prod2];
    \node [nodePre, right = 6.3 cm of prod, name=pre7];
    \node [nodeProd, left = 2cm of pre7, name=prod3];
    \node [nodePre, right = 7.8 cm of prod, name=pre8];
    \node [nodePre, below = 1.7 cm of prod3, name=pre9];
    \node [nodePre, below = 0.8 cm of prod2, name=pre10];
    \node [nodeSum, right = 1.1 cm of pre10, name=sum];
    \node [nodePre, below right = 0.6 cm and -0.84cm of amod, name=pre11];
    \node [nodeSum, below right = 0.78cm and 0.2cm of pre11, name=sum2];
    \node [nodePre, right = 0.4 cm of sum2, name=pre12];
    \node [nodeProd, right = 0.4cm of pre12, name=prod4];
    \node [nodePre, right = 0.4 cm of prod4, name=pre13];
    \node [nodeSum, right = 0.7 cm of pre13, name=sum3];        

    \draw[color=red,thick,dashed] ($(pre.north)+(0,0)$) -- node[solid,pos=1,below,circle,color=red,draw,fill=red,scale=0.4] {}($(amod.south)+(-2.75,0)$) node[pos=0.87,left]{$v_\mathrm{a}$};
    
    \draw[color=DPgreen,thick] ($(pre.west)+(0,0)$) -- ($(veh.east)+(0,0)$) node[pos=1,right,circle,color=DPgreen,draw,fill=DPgreen,scale=0.4] {};
    
    \draw[color=red,thick,dashed] ($(pre2.north)+(0,0)$) -- ($(amod.south)+(1.75,0)$) node[pos=0.87,right]{$n_\mathrm{s,a}$}  node[solid,pos=1,below,circle,color=red,draw,fill=red,scale=0.4]{};
    
    \draw[color=DPgreen,thick] ($(pre2.east)+(0,0)$) -- ($(sub.west)+(0,0)$) node[pos=1,left,circle,color=DPgreen,draw,fill=DPgreen,scale=0.4] {};
    
    \draw[color=red,thick,dashed] ($(pre3.north)+(0,0)$) -- ($(amod.south)+(-1.75,0)$)node[pos=0.87,left]{$s_\mathrm{v,tot}$}  node[solid,pos=1,below,circle,color=red,draw,fill=red,scale=0.4] {};
    
    \draw[color=DPgreen,thick] ($(pre3.south)+(0,0)$) -- ($(prod.north)+(0,0)$) node[pos=1,above,circle,color=DPgreen,draw,fill=DPgreen,scale=0.4,solid] {};
    
    \draw[color=red,thick,dashed] ($(prod.south)+(0,0)$) -- ($(prod.south)+(0,-1.5)$) node[pos=0,below,circle,color=red,draw,fill=red,scale=0.4,solid] {} node[solid,nodePre,below,black,thin]{};
    
    \draw[color=DPgreen,thick] ($(prod.south)+(-0.3,-1.8)$) -- ($(prod.south)+(-2.52,-1.8)$)-| ($(prod.south)+(-2.52,-2.55)$)node[pos=1,below,circle,color=DPgreen,draw,fill=DPgreen,scale=0.4]{};
    
    \draw [-, color=red,thick,dashed] ($(veh.south)+(0.71,0)$) -- ($(pre4.north)+(0,0)$) node[solid,pos=0,below,circle,color=red,draw,fill=red,scale=0.4]{} node[pos=0.5,right]{$C_\mathrm{v,o}$};
    
    \draw [-, color=red,thick,dashed] ($(veh.south)+(-0.79,0)$) -- ($(pre5.north)+(0,0)$) node[solid,pos=0,below,circle,color=red,draw,fill=red,scale=0.4]{} node[pos=0.2,right]{$C_\mathrm{v,f}$} ;
    
    \node[rectangle,align=center] at ($(pre5)+(-1.7,-1.5)$) (pre5Label) {co-design  \\ constraint};
    \draw[->,thick]  (pre5Label) to[out=60,in=135] (pre5);
    
    \draw [-, color=DPgreen,thick]  ($(pre4.east)+(0,0)$) --  ($(prod.west)+(0,0)$) node[pos=1,left,circle,color=DPgreen,draw,fill=DPgreen,scale=0.4] {};
    
    \draw [-, color=red,thick,dashed] ($(sub.south)+(0.775,0)$) -- ($(pre8.north)+(0,0)$) node[solid,pos=0,below,circle,color=red,draw,fill=red,scale=0.4]{} node[pos=0.5,right]{$C_\mathrm{s,o}$};
    
    \draw [-, color=red,thick,dashed] ($(sub.south)+(-0.72,0)$) -- ($(pre7.north)+(0,0)$) node[solid,pos=0,below,circle,color=red,draw,fill=red,scale=0.4]{} node[pos=0.5,right]{$C_\mathrm{s,f}$};
    
    \draw[color=red,thick,dashed] ($(pre6.north)+(0,0)$) -- ($(amod.south)+(-0.75,0)$) node[pos=0.87,right]{$n_\mathrm{v,max}$}  node[solid,pos=1,below,circle,color=red,draw,fill=red,scale=0.4] {};
    
    \draw[color=DPgreen,thick] ($(pre5.east)+(0,0)$) -- ($(prod2.west)+(0,0)$) node[pos=1,left,circle,color=DPgreen,draw,fill=DPgreen,scale=0.4] {};
    \draw[color=DPgreen,thick] ($(pre6.south)+(0,0)$) -- ($(prod2.north)+(0,0)$) node[pos=1,above,circle,color=DPgreen,draw,fill=DPgreen,scale=0.4] {};
    \draw[color=DPgreen,thick] ($(pre2.east)+(0.52,0)$) -- ($(prod3.north)+(0,0)$) node[pos=1,above,circle,color=DPgreen,draw,fill=DPgreen,scale=0.4] {};
    \draw[color=DPgreen,thick] ($(prod3.east)+(0,0)$) -- ($(pre7.west)+(0,0)$) {} node[solid,pos=0,right,circle,color=DPgreen,draw,fill=DPgreen,scale=0.4]{};
    
    \draw[color=red,thick,dashed] ($(prod2.south)+(0,0)$) -- ($(pre10.north)+(0,0)$) node[pos=0,below,circle,color=red,draw,fill=red,scale=0.4,solid] {};
    \draw[color=red,thick,dashed] ($(prod3.south)+(0,0)$) -- ($(pre9.north)+(0,0)$) node[pos=0,below,circle,color=red,draw,fill=red,scale=0.4,solid] {};
    
    \draw[color=DPgreen,thick] ($(pre10.east)+(0,0)$) -- ($(sum.west)+(0,0)$) node[pos=1,left,circle,color=DPgreen,draw,fill=DPgreen,scale=0.4] {};
    \draw[color=DPgreen,thick] ($(pre9.west)+(0,0)$) -- ($(sum.east)+(0,0)$) node[pos=1,right,circle,color=DPgreen,draw,fill=DPgreen,scale=0.4] {};
        
    \draw[color=DPgreen,thick] ($(pre8.south)+(0,0)$) -- ($(pre8.south)+(0,-3)$)|- ($(pre8.south)+(-10.6,-3)$) node[nodeSum,left,black,thin]{} node[pos=1,right,circle,color=DPgreen,draw,fill=DPgreen,scale=0.4] {};
    
    \draw[color=red,thick,dashed] ($(pre8.south)+(-10.92,-3.28)$) -- ($(pre8.south)+(-10.92,-4)$) node[solid,pos=0,below,circle,color=red,draw,fill=red,scale=0.4]{} |-($(pre8.south)+(-10.37,-4)$) node[solid,nodePre,right,black,thin] {};
    
    \draw[color=DPgreen,thick] ($(pre8.south)+(-9.8,-4)$) -- ($(pre8.south)+(-8.5,-4)$)  node[nodeSum,right,black,thin] {}node[pos=1,left,circle,color=DPgreen,draw,fill=DPgreen,scale=0.4] {};
    
    \draw[color=red,thick,dashed]  ($(pre8.south)+(-8.19,-4.29)$) -- ($(pre8.south)+(-8.19,-5)$)node[solid,pos=0,below,circle,color=red,draw,fill=red,scale=0.4]{} node[pos=0.7,right]{$C_\mathrm{tot}$};

    \draw[color=red,thick,dashed] ($(sum.south)+(0,0)$) -- ($(sum.south)+(0,-0.65)$) {}node[pos=0,below,circle,color=red,draw,fill=red,scale=0.4,solid] {};
    \draw[color=red,thick,dashed] ($(sum.south)+(0,-0.9)$) -- ($(sum.south)+(0,-1.45)$) node[nodePre,below,black,thin,solid]{} ;
    
    \draw[color=DPgreen,thick] ($(sum.south)+(-0.29,-1.75)$) -- ($(sum.south)+(-2.18,-1.75)$) {}node[pos=1,right,circle,color=DPgreen,draw,fill=DPgreen,scale=0.4] {};

    \draw[color=DPgreen,thick] ($(amod.north)+(0,1.25)$) -- ($(amod.north)+(0,0)$) node[pos=0.2,right]{$\alpha_\mathrm{tot}$}  node[pos=1,above,circle,color=DPgreen,draw,fill=DPgreen,scale=0.4] {};
    
    \draw[color=DPgreen,thick] ($(pre2.east)+(1.42,0)$) -- ($(pre2.east)+(1.42,0.8)$) {}node[solid,pos=1,below,circle,color=DPgreen,draw,fill=DPgreen,scale=0.4]{};
    
    \draw[color=red,thick,dashed] ($(pre11.north)+(0,0)$) -- ($(amod.south)+(2.75,0)$) node[pos=0.5,right]{$m_\mathrm{CO_2,v,tot}$}  node[solid,pos=1,below,circle,color=red,draw,fill=red,scale=0.4] {};
    
    \draw[color=DPgreen,thick] ($(sum2.west)+(-0.75,0)$) -- ($(sum2.west)+(0,0)$) node[pos=1,left,circle,color=DPgreen,draw,fill=DPgreen,scale=0.4] {} node[pos=0.3,below]{$n_\mathrm{s}$};
    
    \draw[color=DPgreen,thick] ($(prod4.north)+(0,0.5)$) -- ($(prod4.north)+(0,0)$) node[pos=1,above,circle,color=DPgreen,draw,fill=DPgreen,scale=0.4] {}node[pos=0.3,right]{$m_\mathrm{CO_2,s}$};
    \draw[color=red,thick,dashed] ($(sum2.east)+(0,0)$) -- ($(pre12.west)+(0,0)$) node[pos=0,right,circle,color=red,draw,fill=red,scale=0.4,solid] {};
    
    \draw[color=DPgreen,thick] ($(pre12.east)+(0,0)$) -- ($(prod4.west)+(0,0)$) node[pos=1,left,circle,color=DPgreen,draw,fill=DPgreen,scale=0.4] {};
    \draw[color=red,thick,dashed] ($(prod4.east)+(0,0)$) -- ($(pre13.west)+(0,0)$) node[pos=0,right,circle,color=red,draw,fill=red,scale=0.4,solid] {};
    \draw[color=DPgreen,thick] ($(pre13.east)+(0,0)$) -- ($(sum3.west)+(0,0)$) node[pos=1,left,circle,color=DPgreen,draw,fill=DPgreen,scale=0.4] {};
    
    \draw[color=DPgreen,thick] ($(pre11.east)+(0,0)$) -- ($(sum3.north)+(0,1)$)-| ($(sum3.north)+(0,0)$)node[pos=1,above,circle,color=DPgreen,draw,fill=DPgreen,scale=0.4] {};
    
    \draw[color=red,thick,dashed] ($(sum3.south)+(0,0)$) -- ($(sum3.south)+(0,-6.75)$) node[pos=0,below,circle,color=red,draw,fill=red,scale=0.4,solid] {} |- ($(sum3.south)+(-4,-6.75)$) -|($(sum3.south)+(-4,-7.73)$)node[pos=0.9,right]{$m_\mathrm{CO_2,tot}$};
    
    \draw[color=red, thick,dashed] ($(amod.south)+(0.75,0)$) -- node[solid,pos=0,below,circle,color=red,draw,fill=red,scale=0.4] {} ($(amod.south)+(0.75,-5)$) |- ($(amod.south)+(1.75,-5)$)-|($(amod.south)+(1.75,-6.85)$){};
    
    \draw[color=red, thick,dashed]  ($(amod.south)+(1.75,-7.1)$) -- ($(amod.south)+(1.75,-7.88)$) {};
    \draw[color=red, thick,dashed]  ($(amod.south)+(1.75,-8.08)$) -- ($(amod.south)+(1.75,-9.95)$) node[pos=0.92,right]{$t_\mathrm{avg}$};
    
    \draw[color=DPgreen,thick] ($(prod2.east)+(0.75,0)$) -- ($(prod2.east)+(0,0)$) node[pos=1,right,circle,color=DPgreen,draw,fill=DPgreen,scale=0.4] {} node[pos=0.5,above,DPgreen]{$\frac{1}{l_\mathrm{v}}$};
    
    \draw[color=DPgreen,thick] ($(prod3.west)+(-0.75,0)$) -- ($(prod3.west)+(0,0)$) node[pos=1,left,circle,color=DPgreen,draw,fill=DPgreen,scale=0.4] {} node[pos=0.5,above,DPgreen]{$\frac{1}{l_\mathrm{s}}$};
    
    \node[rectangle,align=center,color=red] at ($(1.9,-11)$) (name) {total cost};
    \node[rectangle,align=center,color=red] at ($(5.1,-11.3)$) (name2) {average \\ travel time};
    \node[rectangle,align=center,color=red] at ($(7.2,-11.25)$) (name3) {total \\ emissions};
    \node[rectangle,align=center,color=DPgreen] at ($(3.4,2.5)$) (name3) {total \\ request rate};
    
    \draw[dashed] (-3.9,-10.2) -- (12,-10.2);
    \draw[dashed] (-3.9,-10.2) -- (-3.9,1.2);
    \draw[dashed] (-3.9,1.2) -- (12,1.2);
    \draw[dashed] (12,1.2) -- (12,-10.2);
    
    \end{tikzpicture}}
    \end{center}
    \caption{Co-design problem of the full system.}
    \label{fig:mcdp}
    \end{subfigure}   
    \caption{Schematic representation of the individual design problems (a-c) as well as of the co-design problem of the full system (d). In solid \textcolor{DPgreen}{green} the provided functionalities and in dashed \textcolor{red}{red} the required resources. The edges in the co-design diagram (d) represent co-design constraints: The resources required by a first design problem are the lower bound for the functionalities provided by the second one.}
    \end{center}
\end{figure}
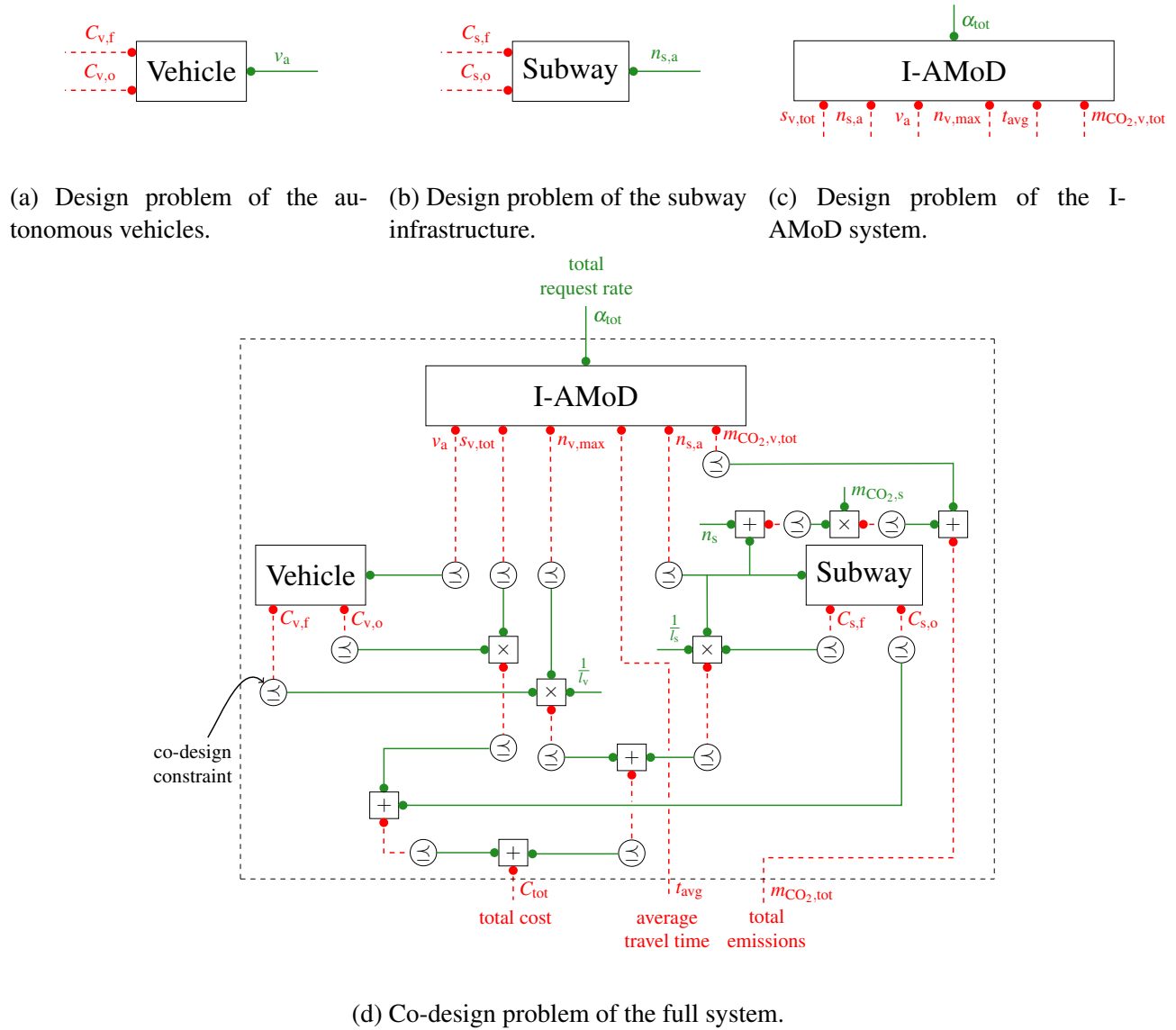

The \gls{abk:iamod} DP (see Figure \ref{fig:amoddp}) provides the demand satisfaction as a functionality, expressed through the total customer request rate
\begin{linenomath}
\begin{equation}
\label{eq:requests}
\alpha_\mathrm{tot}\coloneqq\sum_{m\in \mathcal{M}} \alpha_m.
\end{equation}
\end{linenomath}
To successfully satisfy a given set of travel requests, we require the following resources:  (i) the achievable speed of the \glspl{abk:av} $v_\mathrm{a}$, (ii) the number of available \glspl{abk:av} per fleet $n_\mathrm{v,max}$, (iii) the number of trains $n_\mathrm{s,a}$ acquired by the public transportation system, and (iv) the average travel time of a trip
\begin{linenomath}
\begin{equation}
    \label{eq:traveltime}
    t_\mathrm{avg}\coloneqq\frac{1}{\alpha_\mathrm{tot}}\cdot \sum_{m\in \mathcal{M}, (i,j)\in \mathcal{A}} t_{ij} \cdot f_m(i,j),
\end{equation}
\end{linenomath}
(v) the total distance driven by the \glspl{abk:av} per unit time
\begin{linenomath}
\begin{equation}
    s_\mathrm{v,tot}\coloneqq\sum_{(i,j)\in \mathcal{A}_\mathrm{R}}s_{ij} \cdot \left(f_0(i,j)+\sum_{m\in \mathcal{M}}f_m(i,j)\right),
\end{equation}
\end{linenomath}
and (vi) the total \glspl{abk:av} CO\textsubscript{2} emissions per unit time
\begin{linenomath}
    \begin{equation}
    m_\mathrm{CO_2,v,tot}\coloneqq\gamma \cdot \sum_{(i,j)\in \mathcal{A}_\mathrm{R}} e_{ij} \cdot \left(f_0(i,j)+\sum_{m\in \mathcal{M}}f_m(i,j)\right),
\end{equation}
\end{linenomath}
where $\gamma$ relates the energy consumption and the CO\textsubscript{2} emissions. We assume that customers trips and \gls{abk:amod} rebalancing strategies are chosen to maximize the customers welfare, defined through the average travel time $t_\mathrm{avg}$. Hence, we link the functionality and resources of the \gls{abk:iamod} DP through the following optimization problem:
\begin{linenomath}
\begin{equation}
\label{eq:TIamodopt}
\min_{\{f_m(\cdot,\cdot)\}_m, f_0(\cdot,\cdot)}t_\mathrm{avg}=\frac{1}{\alpha_\mathrm{tot}}\sum_{m\in \mathcal{M}, (i,j)\in \mathcal{A}} t_{ij} \cdot f_m(i,j)
\quad
\mathrm{ s.t. \ Eq. }\,\eqref{eq:flowconstotal},
\mathrm{ Eq. } \,\eqref{eq:capacity},
\mathrm{ Eq. } \,\eqref{eq:fleetsize}.
\end{equation}
\end{linenomath}
Formally, $\mathcal{F}_\mathrm{o}=\overline{\mathbb{R}}_+$, and $\mathcal{R}_\mathrm{o}=\overline{\mathbb{R}}_+\times \overline{\mathbb{N}} \times \overline{\mathbb{N}} \times \overline{\mathbb{R}}_+ \times \overline{\mathbb{R}}_+ \times
\overline{\mathbb{R}}_+$.
Note that in general, the optimization problem \eqref{eq:TIamodopt} might possess multiple optimal solutions, making the relation between resources and functionality ill-posed. To overcome this subtlety, if two solutions share the same average travel time, we select the one incurring in the \emph{lowest} mileage.

\subsubsection{The Monotone Co-Design Problem}
\label{sec:mcdp}
The full-system \gls{abk:cdp} results from the interconnection of the \glspl{abk:dp} presented above. A schematic representation is shown in Figure~\ref{fig:mcdp}.
The functionality of the system is to provide mobility service to the customers, quantified, as in~\eqref{eq:requests}, by means of the total request rate.
To this end, the following three resources are required.
First, on the customers side, we require an average travel time, defined as in~\eqref{eq:traveltime}.
Second, on the central authority side, the resource is the total transportation cost of the intermodal mobility system: Assuming an average vehicles' life of $l_\mathrm{v}$, an average trains' life of $l_\mathrm{s}$, and a baseline subway fleet of $n_\mathrm{s,baseline}$ trains, we express the total costs as
\begin{linenomath}
\begin{equation}
C_\mathrm{tot}=C_\mathrm{v}+C_\mathrm{s},
\end{equation}
\end{linenomath}
where $C_\mathrm{v}$ is the \glspl{abk:av}-related cost
\begin{linenomath}
\begin{equation}
C_\mathrm{v}=\frac{C_\mathrm{v,f}}{l_\mathrm{v}}\cdot n_\mathrm{v} + C_\mathrm{v,o}\cdot s_\mathrm{v,tot},
\end{equation}
\end{linenomath}
and $C_\mathrm{s}$ is the public transit-related cost
\begin{linenomath}
\begin{equation}
C_\mathrm{s}=\frac{C_\mathrm{s,f}}{l_\mathrm{s}}\cdot n_\mathrm{s,a} + C_\mathrm{s,o}.
\end{equation}
\end{linenomath}
Third, on the environmental side, the resources is the total CO\textsubscript{2} emissions
\begin{linenomath}
\begin{equation}
m_\mathrm{CO_2,tot}=m_\mathrm{CO_2,v,tot}+m_\mathrm{CO_2,s}\cdot n_\mathrm{s},
\end{equation}
\end{linenomath}
where $m_\mathrm{CO_2,s}$ represents the CO\textsubscript{2} emissions of a single train.
Formally, $\alpha_\mathrm{tot}$ is the \gls{abk:cdp} functionality, whereas $t_\mathrm{avg}$, $C_\mathrm{tot}$, and $m_\mathrm{CO_2,tot}$ are the resources. Consistently, the functionality space is $\mathcal{F}=\overline{\mathbb{R}}_+$ and the resources space is $\mathcal{R}=\overline{\mathbb{R}}_+\times \overline{\mathbb{R}}_+\times \overline{\mathbb{R}}_+$. Note that the resulting \gls{abk:cdp} is indeed monotone, since it consists of the interconnection of monotone \glspl{abk:dp}~\cite{Censi2015}.

\subsection{Discussion}
A few comments are in order.
First, we lump the vehicle autonomy in its achievable velocity. We leave to future research more elaborated \gls{abk:av} models, accounting for instance for accidents rates~\cite{Richards2010} and for safety levels. 
Second, we assume the service frequency of the subway system to scale linearly with the number of trains. We inherently rely on the assumption that the existing infrastructure can homogeneously accommodate the acquired train cars. To justify the assumption, we include an upper bound on the number of potentially acquirable trains in our case study design in Section~\ref{sec:results}.
Third, we highlight that the \gls{abk:iamod} framework is only one of the many feasible ways to map total demand to travel time, costs, and emissions. Specifically, practitioners can easily replace the corresponding \gls{abk:dp} with more sophisticated models (e.g., simulation-based frameworks like MATSim~\cite{HorniAxhausen2016}), as long as the monotonicity of the system is preserved. In our setting, we conjecture the customers and vehicles routes to be centrally controlled by the central authority in a socially-optimal fashion.
Fourth, we assume a homogenous fleet of \glspl{abk:av}. Nevertheless, our model is readily extendable to capture heterogeneous fleets. 
Finally, we consider a fixed travel demand, and compute the antichain of resources providing it. Nonetheless, our formalization can be readily extended to arbitrary demand models preserving the monotonicity of the \gls{abk:cdp} (accounting for instance for elastic and stochastic effects). We leave this topic to future research.

\section{Results}\label{sec:results}
In this section, we leverage the framework presented in Section~\ref{sec:model} to evaluate the real-world case of Washington D.C., USA. Section \ref{sec:case1} details the case study. We then present numerical results in Sections~\ref{sec:case2} and \ref{sec:case3}.

\subsection{Case Study}
\label{sec:case1}
We present our studies on the real-world case of Washington D.C., USA.
We import the road network (Figure~\ref{fig:dc}) and its features from OpenStreetMap~ \cite{HaklayWeber2008}. 
The public transit network and its schedules are extracted from the GTFS data~\cite{GTFS2019}. The travel demand is obtained by unifying the origin-destination pairs of the morning peak of May 31st 2017 provided by the taxi companies~\cite{ODDC2017} and the Washington Metropolitan Area Transit Authority (WMATA)~\cite{PIM2012}. Given the lack of reliable demand data for the MetroBus system, we focus our studies on the MetroRail system and its design, inherently assuming  MetroBus commuters to be unaffected by our design methodology. To account for the large presence of ride-hailing companies, we scale the taxi demand rate by a factor of 5~\cite{Siddiqui2018b}. Overall, the demand dataset includes 15,872 travel requests, corresponding to a demand rate of \unitfrac[24.22]{requests}{s}.
To account for congestion effects, we compute the nominal road capacity as in~\cite{DoA1977} and assume an average baseline road usage of 93\%, in line with~\cite{DixonIrshadEtAl2018}. We summarize the main parameters together with their bibliographic sources in Table \ref{tab:params}. In the remainder of this section, we tailor and solve the co-design problem presented in Section \ref{sec:model} through the PyMCDP solver~\cite{Censi2019}, and investigate the influence of different \glspl{abk:av} costs on the design objectives and strategies.

\begin{table}[h]
	\begin{center}
		\begin{scriptsize}
		\begin{tabular}{lllccccclc}
			\toprule
			\multicolumn{2}{l}{\textbf{Parameter}} & \textbf{Variable}  & \multicolumn{5}{c}{\textbf{Value}} &\textbf{Units}& \textbf{Source}\\
			Baseline road usage & & $u_{ij}$ &\multicolumn{5}{c}{93} &\unit[]{\%}& ~\cite{DixonIrshadEtAl2018}\\
			\midrule
			&&& \textbf{Case 1} & \textbf{Case 2.1} & \textbf{Case 2.2}&\textbf{Case 3.1}&\textbf{Case 3.2}\\ \cline{4-8}  \\[-1.0em]
			\multicolumn{2}{l}{Vehicle operational cost} & $C_\mathrm{v,o}$ &0.084 & 0.084&0.062 &0.084&0.084&\unitfrac[]{USD}{mile}&~\cite{PavlenkoSlowikEtAl2019, BoeschBeckerEtAl2018}\\
			\multicolumn{2}{l}{Vehicle cost} & $C_\mathrm{v,v}$ & 32,000&32,000&26,000&32,000&32,000&\unitfrac[]{USD}{car}& ~\cite{PavlenkoSlowikEtAl2019}\\
			\multirow{7}{*}{Vehicle automation cost} & \unit[20]{mph} & \multirow{7}{*}{$C_\mathrm{v,a}$} &15,000&20,000&3,700&0&500,000&\unitfrac[]{USD}{car}& ~\cite{BoeschBeckerEtAl2018,FagnantKockelman2015,BauerGreenblattEtAl2018,Litman2019,Wadud2017}\\
			 & \unit[25]{mph} &  &15,000&30,000&4,400&0&500,000& \unitfrac[]{USD}{car}&~\cite{BoeschBeckerEtAl2018,FagnantKockelman2015,BauerGreenblattEtAl2018,Litman2019,Wadud2017}\\
			 & \unit[30]{mph} &  &15,000&55,000 &6,200&0&500,000&\unitfrac[]{USD}{car}& ~\cite{BoeschBeckerEtAl2018,FagnantKockelman2015,BauerGreenblattEtAl2018,Litman2019,Wadud2017}\\
			 & \unit[35]{mph} &  &15,000&90,000&8,700&0&500,000&\unitfrac[]{USD}{car}& ~\cite{BoeschBeckerEtAl2018,FagnantKockelman2015,BauerGreenblattEtAl2018,Litman2019,Wadud2017}\\
			 & \unit[40]{mph} &  &15,000&115,000 &9,800&0&500,000& \unitfrac[]{USD}{car}&~\cite{BoeschBeckerEtAl2018,FagnantKockelman2015,BauerGreenblattEtAl2018,Litman2019,Wadud2017}\\
			 & \unit[45]{mph} &  &15,000&130,000 &12,000&0&500,000& \unitfrac[]{USD}{car}&~\cite{BoeschBeckerEtAl2018,FagnantKockelman2015,BauerGreenblattEtAl2018,Litman2019,Wadud2017}\\
			 & \unit[50]{mph} &  &15,000&150,000 &13,000&0&500,000&\unitfrac[]{USD}{car}& ~\cite{BoeschBeckerEtAl2018,FagnantKockelman2015,BauerGreenblattEtAl2018,Litman2019,Wadud2017}\\
			\multicolumn{2}{l}{Vehicle life} &$l_\mathrm{v}$ &5 & 5& 5 &5&5&\unit[]{years}&~\cite{PavlenkoSlowikEtAl2019}\\
			\multicolumn{2}{l}{CO$_2$ per Joule}& $\gamma$ & 0.14&0.14&0.14&0.14&0.14&\unitfrac[]{g}{kJ}& ~\cite{Watttime2018}\\
			\multicolumn{2}{l}{Time from $\mathcal{G}_\mathrm{W}$ to $\mathcal{G}_\mathrm{R}$ } & $t_\mathrm{WR}$ & 300&300&300&300&300&\unit[]{s}&-\\
			\multicolumn{2}{l}{Time from $\mathcal{G}_\mathrm{R}$ to $\mathcal{G}_\mathrm{W}$} & $t_\mathrm{RW}$ & 60&60&60&60&60&\unit[]{s}&-\\
			\multicolumn{2}{l}{Speed limit fraction}&$\beta$&$\frac{1}{1.3}$&$\frac{1}{1.3}$&$\frac{1}{1.3}$&$\frac{1}{1.3}$&$\frac{1}{1.3}$&\unit[]{-}&~\cite{Dahl2018}\\
			\midrule
			\multirow{3}{*}{Subway operational cost} & \unit[100]{\%}& \multirow{3}{*}{$C_\mathrm{s,o}$} & \multicolumn{5}{c}{148,000,000} &\unitfrac[]{USD}{year}&~\cite{WMATA2017}\\
			& \unit[133]{\%}&  & \multicolumn{5}{c}{197,000,000} &\unitfrac[]{USD}{year}& ~\cite{WMATA2017}\\
		    & \unit[200]{\%}&  & \multicolumn{5}{c}{295,000,000} &\unitfrac[]{USD}{year}& ~\cite{WMATA2017}\\			
			\multicolumn{2}{l}{Subway fixed cost} & $C_\mathrm{s,f}$ &  \multicolumn{5}{c}{14,500,000}  &\unitfrac[]{USD}{train}& ~\cite{Aratani2015}\\
			\multicolumn{2}{l}{Train life}& $l_\mathrm{s}$ & \multicolumn{5}{c}{30} &{\unit[]{years}}& ~\cite{Aratani2015}\\ 
			\multicolumn{2}{l}{Subway CO$_2$ emissions per train}& $m_\mathrm{CO_2,s}$ &\multicolumn{5}{c}{140} &\unitfrac[]{ton}{year}&~\cite{WMATA2018} \\
			\multicolumn{2}{l}{Train fleet baseline} & $n_\mathrm{s,baseline}$ &\multicolumn{5}{c}{112}&\unit[]{trains}& ~\cite{Aratani2015}\\
			\multicolumn{2}{l}{Subway service frequency}& $\varphi_{j,\mathrm{baseline}}$ & \multicolumn{5}{c}{$\frac{1}{6}$}&\unitfrac[]{1}{minutes}& ~\cite{Jaffe2015}\\
			\multicolumn{2}{l}{Time from $\mathcal{G}_\mathrm{W}$ to $\mathcal{G}_\mathrm{P}$ and vice-versa} & $t_\mathrm{WS}$ & \multicolumn{5}{c}{$60$}&\unit[]{s}&-\\
		\bottomrule
	\end{tabular}
	\caption{Parameters, variables, numbers, and units for the case studies.}
	\label{tab:params}
\end{scriptsize}
\end{center}
\end{table}

\subsection{Case 1 - Constant Cost of Automation}
\label{sec:case2}
\begin{figure}[!htb]
    \begin{center}
    \begin{subfigure}[h]{\textwidth}
    \begin{center}
    \includegraphics[width=\textwidth]{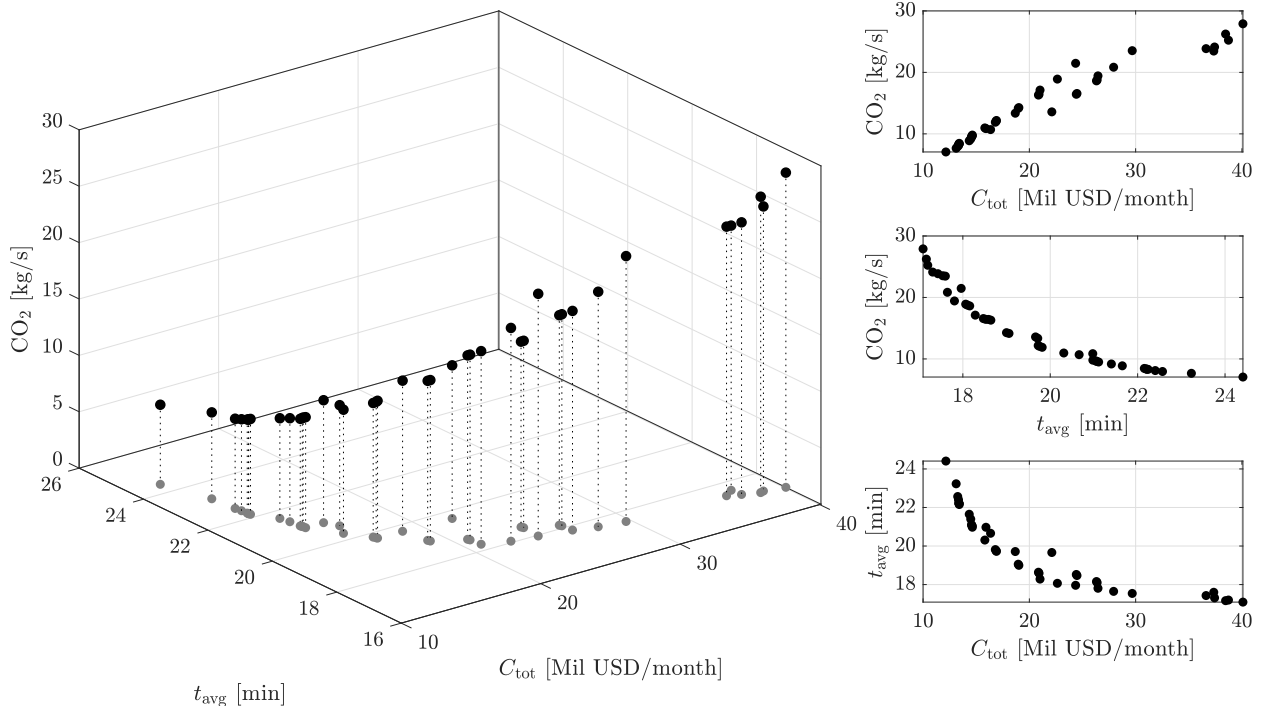}
    \caption{Tri-dimensional representation antichain elements and their projection in the cost-time space (left) and their two-dimensional projections (right).}
    \label{fig:resultsart3D}
    \end{center}
    \end{subfigure}
    ~
    \vspace{0.5cm}
    \begin{subfigure}[h]{\textwidth}
    \begin{center}
\includegraphics[width=\textwidth]{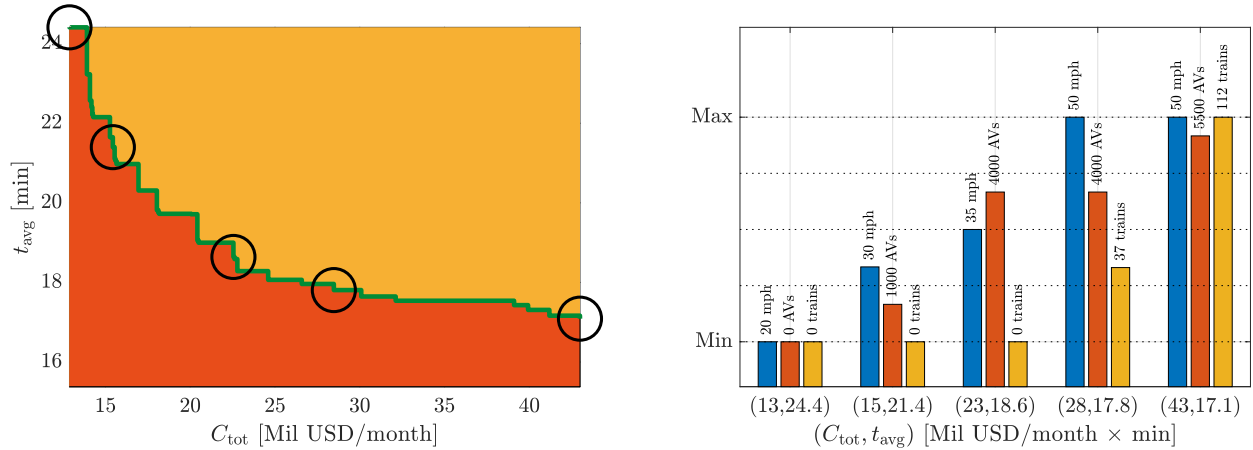}
    \end{center}
    \caption{Results for constant automation costs. On the left, the two-dimensional representation of the antichain elements: in red are the unfeasible strategies, in orange the feasible but irrational solutions, and in green the Pareto front. On the right, the implementations corresponding to the highlighted antichain elements, quantified in terms of achievable vehicle speed, \glspl{abk:av} fleet size, and train fleet size.}
    \label{fig:resultsart2D}
    \end{subfigure}
    \caption{Solution of the Co-Design Problem (CDP) for the state-of-the art case.}
    \label{fig:resultsart}
    \end{center}
\end{figure}

In line with~\cite{BoeschBeckerEtAl2018,FagnantKockelman2015,BauerGreenblattEtAl2018,Litman2019,Wadud2017}, we first assume an average achievable-velocity-independent cost of automation. As discussed in Section \ref{sec:model}, we design the system by means of subway service frequency, \glspl{abk:av} fleet size, and achievable free-flow speed. Specifically, we allow the municipality to (i) increase the subway service frequency $\varphi_j$ by a factor of 0\%, 33\%, or 100\%, (ii) deploy an \gls{abk:amod} fleet of size $n_\mathrm{v,max} \in \{0,500,1000,\hdots,6000\}$ vehicles, and (iii) design the single \gls{abk:av} achievable velocity $v_\mathrm{a} \in \{\unit[20]{mph},\unit[25]{mph},\hdots,\unit[50]{mph}\}$. 
We assume the \gls{abk:amod} fleet to be composed of battery electric BEV-250 mile vehicles~\cite{PavlenkoSlowikEtAl2019}.
In Figure~\ref{fig:resultsart3D}, we show the solution of the co-design problem by reporting the antichain consisting of the total transportation cost, average travel time, and total CO\textsubscript{2} emissions. These solutions are \emph{rational} (and not comparable) in the sense that there exists no instance which simultaneously yields lower cost, average travel time, and emissions. 
For the sake of clarity, we opt for a two-dimensional antichain representation, by translating and including the emissions in the total cost. To do so, we consider the conversion factor \unitfrac[40]{USD}{kg}~\cite{HowardSylvan2015}. Note that since this transformation preserves the monotonicity of the \gls{abk:cdp} it smoothly integrates in our framework.
Doing so, we can conveniently depict the co-design strategies through the two-dimensional antichain (Figure~\ref{fig:resultsart2D}, right) and the corresponding municipality actions (Figure~\ref{fig:resultsart2D}, left).
Generally, as the municipality budget increases, the average travel time per trip required to satisfy the given demand decreases, reaching a minimum of about \unit[17.1]{minutes} with a monthly expense of around \unitfrac[43,000,000]{USD}{month}.
This configuration corresponds to a fleet \unit[5,500]{AVs} able to drive at \unit[50]{mph} and to the doubling of the current MetroRail train fleet.
On the other hand, the smallest rational investment of \unitfrac[12,900,000]{USD}{month} leads to a \unit[42]{\%} higher average travel time, corresponding to a non-existent autonomous fleet and an unchanged subway infrastructure. Notably, an expense of \unitfrac[23,000,000]{USD}{month} (\unit[48]{\%} lower than the highest rational investment) only increases the minimal required travel time by \unit[9]{\%}, requiring a fleet of \unit[4,000]{vehicles} able to drive at \unit[35]{mph} and no acquisition of trains. Conversely, an investment of \unitfrac[15,600,000]{USD}{month} (just \unitfrac[2,700,000]{USD}{month} more than the minimal rational investment) provides a \unit[3]{min} shorter travel time. 
Remarkably, the design of \glspl{abk:av} able to exceed \unit[40]{mph} only improves the average travel time by \unit[6]{\%}, and it is rational just starting from an expense of \unitfrac[22,800,000]{USD}{month}.
This suggests that the design of faster vehicles mainly results in higher emission rates and costs, without substantially contributing to a more time-efficient demand satisfaction. 
Finally, it is rational to improve the subway system only starting from a budget \unitfrac[28,500,000]{USD}{month}, leading to a travel time improvement of just \unit[4]{\%}. This trend can be explained with the high train acquisition and increased operation costs, related to the subway reinforcement. We expect this phenomenon to be more marked for other cities, considering the moderate operation costs of the MetroRail subway system due to its automation~\cite{Jaffe2015} and related benefits~\cite{WangZhangEtAl2016}.

\subsection{Case 2 - Speed-Dependent Automation Costs}
\label{sec:case3}
To relax the potentially unrealistic assumption of a velocity-independent automation cost, we consider a performance-dependent cost structure. The large variance in sensing technologies and their reported performances~\cite{GawronKeoleianEtAl2018} suggests that this rationale is reasonable. Indeed, the technology required today to safely operate an autonomous vehicle at \unit[50]{mph} is substantially more sophisticated, and therefore more expensive, than the one needed at \unit[20]{mph}. To this end, we adopt the cost structure reported in Table \ref{tab:params}. Furthermore, the frenetic evolution of automation techniques intricates their monetary quantification~\cite{WCP2018}. Therefore, we perform our studies with current (2019) costs as well as with their projections for the upcoming decade (2025)~\cite{Lienert2019,PavlenkoSlowikEtAl2019}.

\subsubsection{Case 2.1 - 2019}
\begin{figure}[!htb]
    \begin{center}
    \begin{subfigure}[h]{\textwidth}
    \includegraphics[width=\textwidth]{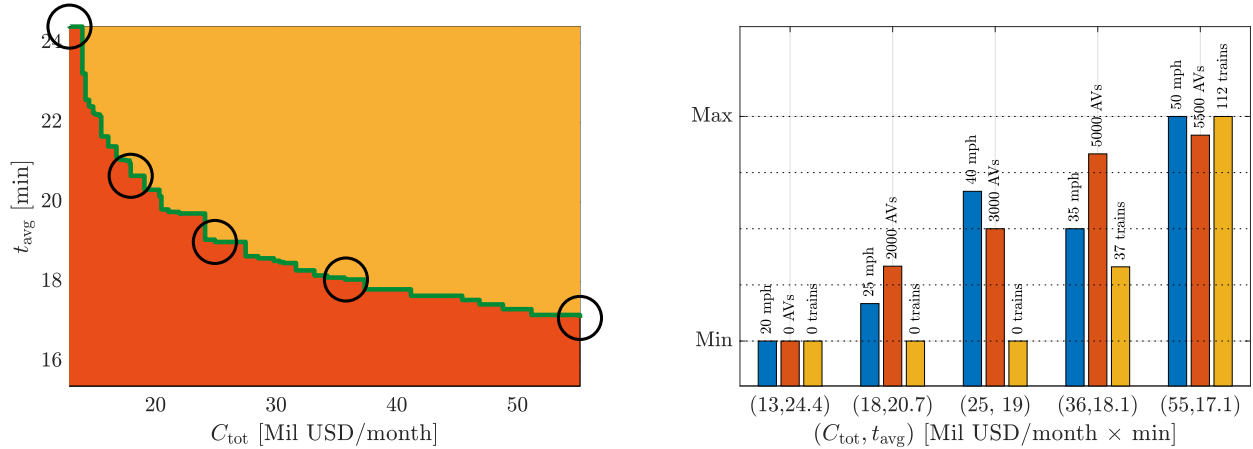}
    \caption{Speed-dependent automation costs in 2019.}
    \label{fig:results19}
     \end{subfigure}
     ~
     \begin{subfigure}[h]{\textwidth}
    \includegraphics[width=\textwidth]{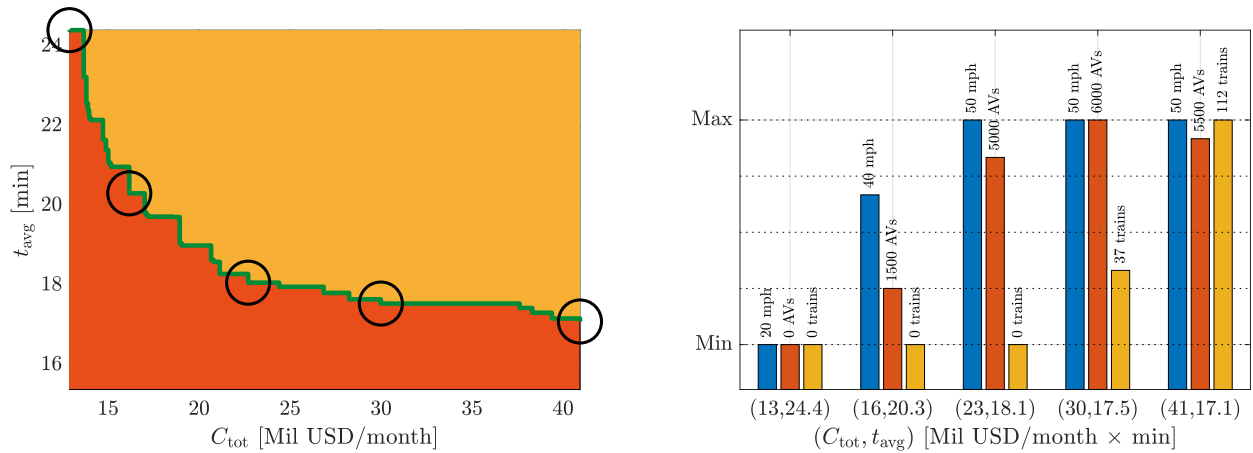}
    \caption{Speed-dependent automation costs in 2025.}
    \label{fig:results25}
     \end{subfigure}
    \end{center}
    \caption{Results for the speed-dependent automation costs. On the left, the two-dimensional representation of the antichain elements: in red are the unfeasible strategies, in orange the feasible but irrational solutions, and in green the Pareto front. On the right, the implementations corresponding to the highlighted antichain elements.}
    \label{fig:results1925}
\end{figure}

Here, we study the hypothetical case of an immediate \glspl{abk:av} fleet deployment, assuming the technological advances to be provided. We introduce the aforementioned velocity-dependent automation cost structure and obtain the results reported in Figure \ref{fig:results19}. Comparing these results with the state-of-the-art values presented in Figure \ref{fig:resultsart} confirms the previously observed trend concerning elevated vehicle speeds. Indeed, spending \unitfrac[24,900,000]{USD}{month} (\unit[55]{\%} lower than the highest rational expense) only increases the average travel time by \unit[10]{\%}, requiring a fleet of \unit[3,000]{\glspl{abk:av}} at \unit[40]{mph} and no subway interventions. Nevertheless, the comparison shows two substantial differences.
First, the budget required to reach the minimum travel time of \unit[17.1]{minutes} is \unit[28]{\%} higher compared to the previous case, and consists of the same strategy for the municipality, i.e., doubling the train fleet and having a fleet of \unit[5,500]{\glspl{abk:av}} at \unit[50]{mph}.
Second, the higher vehicle costs result in an average \glspl{abk:av} fleet growth of \unit[5]{\%}, an average velocity reduction of \unit[9]{\%}, and an average train fleet growth of \unit[7]{\%}. The latter suggests a shift towards a poorer \glspl{abk:av} performance in favour of fleets enlargements.  
\subsubsection{Case 2.2 - 2025}
Experts forecast a large automation cost reduction (up to \unit[90]{\%}) in the next decade, due to mass-production of the \glspl{abk:av} sensing technology~\cite{Lienert2019,WCP2018}.
In line with this vision, we inspect the futuristic scenario by solving the co-design problem for the adapted automation costs, and report the results in Figure \ref{fig:results25}.
Two comments are in order. First, one can notice that the maximal rational budget is \unit[25]{\%} lower than in the immediate adoption case. Second, the reduction in autonomy costs clearly eases the acquisition of more performant \glspl{abk:av}, increasing the average vehicle speed by \unit[10]{\%}. As a direct consequence, the \glspl{abk:av} and train fleets are reduced in size by \unit[5]{\%} and \unit[10]{\%}, respectively. 

\subsection{Case 3 - Asymptotic Automation Cost Analysis}
We conclude our numerical analysis with a study on asymptotic cost structures. Specifically, we compare the constrasting cases of a free-of-charge and a very high (\unitfrac[500,000]{USD}{car}) automation cost.

\subsubsection{Case 3.1- No Automation Cost}
\begin{figure}[!htb]
    \begin{center}
    \begin{subfigure}[h]{\textwidth}
    \includegraphics[width=\textwidth]{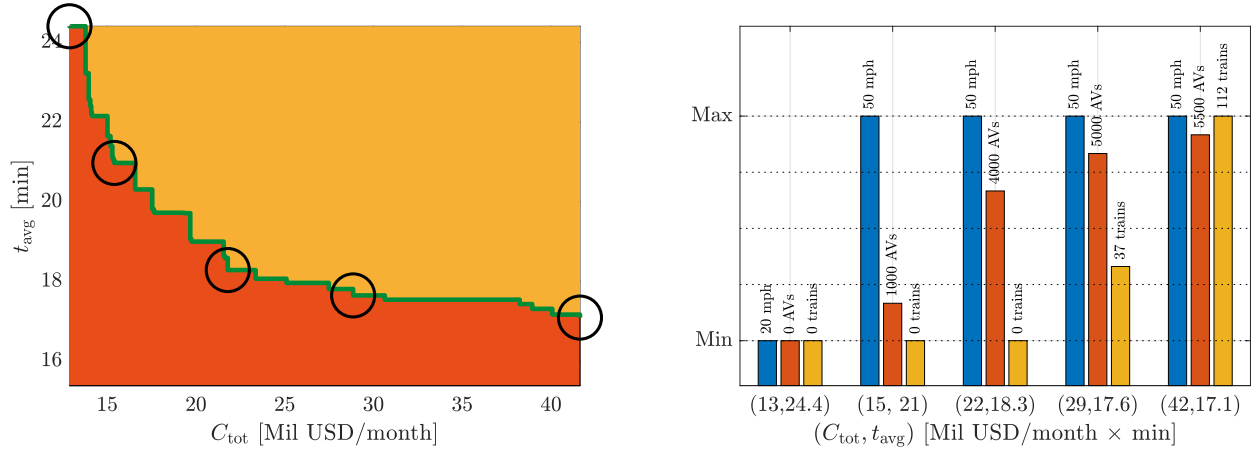}
    \caption{Results for no automation costs.}
    \label{fig:resultslow}
     \end{subfigure}
     ~
     \begin{subfigure}[h]{\textwidth}
    \includegraphics[width=\textwidth]{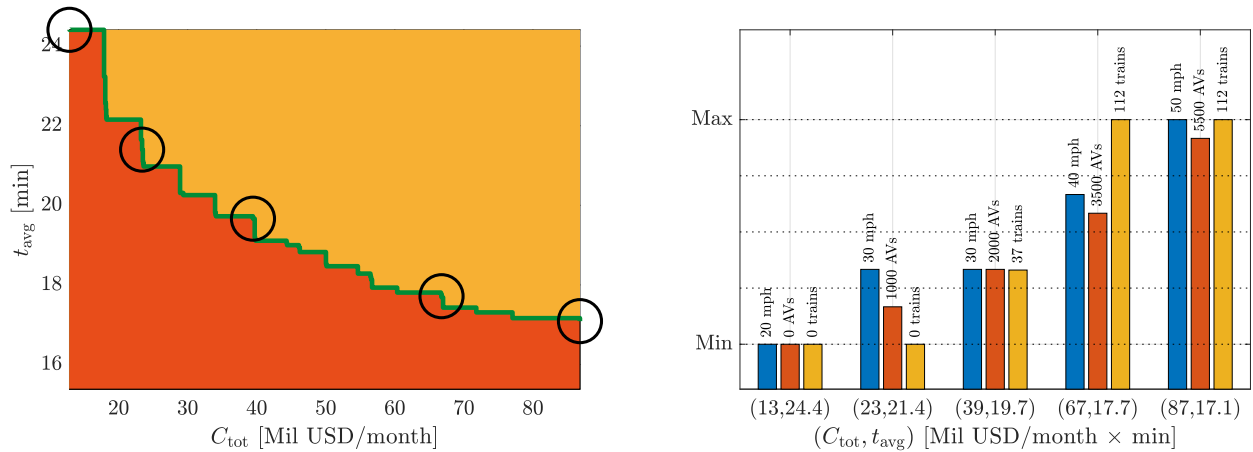}
    \caption{Results for large automation costs.}
    \label{fig:resultshigh}
     \end{subfigure}
    \end{center}
    \caption{Results for asymptotic automation costs. On the left, the two-dimensional representation of the antichain elements: in red are the unfeasible strategies, in orange the feasible but irrational solutions, and in green the Pareto front. On the right, the implementations corresponding to the highlighted antichain elements.}
    \label{fig:resultslowhigh}
\end{figure}

This case could represent the future full deployment of \glspl{abk:av}, where the cost of automation is naturally included in the normal production costs and does not represent an additional expense.
The results for this cost structure are reported in Figure \ref{fig:resultslow}. Notably, the Pareto front is very similar to the one presented in Figure \ref{fig:resultsart}, leading to just a \unit[5]{\%} increase in the \glspl{abk:av} speed and to invariate fleet sizes. Clearly, the assumption a cost-free automation favours the deployment of \glspl{abk:av} at the expenses of subway improvements. 

\subsubsection{Case 3.2- High Automation Cost}
In this case we assume a performance-independent automation cost of \unitfrac[500,000]{USD}{car}. Although this cost may appear unreasonably large, it roughly captures the extremely onerous research and development costs which \glspl{abk:av} companies are facing today~\cite{Korosec2019}. Indeed, no company has shown the ability to safely and reliably deploy large fleets of \glspl{abk:av} yet. The results, reported in Figure~\ref{fig:resultshigh}, show different trends from the ones depicted in Figure~\ref{fig:resultslow}.
First, we observe substantial shift towards an increase of \unit[163]{\%} in the average train fleet size, followed by a \unit[17]{\%} decrease in the average \glspl{abk:av} fleet size.
Second, to reach the minimal rational average travel time, one needs an expense of roughly \unitfrac[87,000,000]{USD}{month}, corresponding to a \unit[107]{\%} higher investment.

\subsection{Discussion}
\begin{figure}[!htb]
    \begin{center}
    \includegraphics[width=0.75\textwidth]{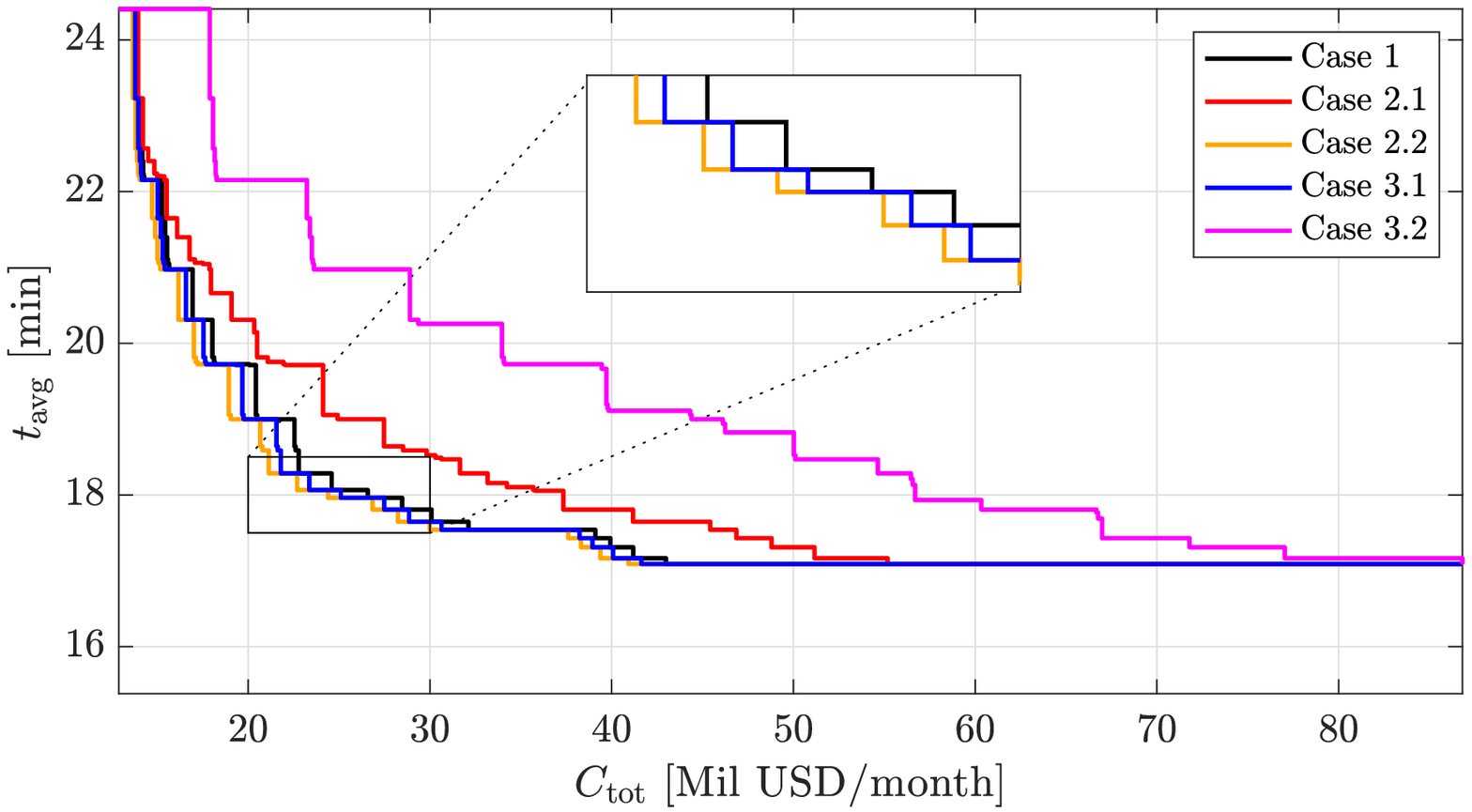}
    \end{center}
    \caption{Comparison of the antichains resulting from different case studies.}
    \label{fig:resultsmix}
\end{figure}

A few comments are in order. First, the presented case studies illustrate the ability of our framework to extract the set of rational design strategies for an \glspl{abk:av}-enabled mobility system.
This way, stakeholders such as \glspl{abk:av} companies, transportation authorities, and policy makers get transparent and interpretable insights on the impact of future interventions. 
Second, we perform a sensitivity analysis (Figure \ref{fig:resultsmix}) through the variation of the autonomy cost structures.
On the one hand, this reveals a clear transition from small fleets of fast \glspl{abk:av} (in the case of low autonomy costs) to a fleet of numerous slow vehicles (in the case of high autonomy costs).
On the other hand, our studies highlight that investments in the public transit infrastructure are rational only when large budgets are available.
Indeed, the onerous train acquisition and operation costs lead to a comparative advantage of \glspl{abk:av}-based mobility.
In the future, we plan to collect more high-resolution data to corroborate our conclusions with quantitative results.

\section{Conclusion}\label{sec:conclusion}
In this paper, we leveraged the mathematical theory of co-design to propose a design framework for \glspl{abk:av}-enabled mobility systems. Specifically, the nature of our framework allows both for the modular and compositional interconnection of the design problems of different mobility options and for multiple objectives.
Starting from the multi-commodity flow model of an intermodal autonomous mobility-on-demand system, we designed autonomous vehicles and public transit both from a vehicle-centric and fleet-level perspective. In particular, we studied the problem of deploying a fleet of self-driving vehicles providing on-demand mobility in cooperation with public transit, adapting the speed achievable by the vehicles, the fleet size, and the service frequency of the subway lines. 
Our framework allows the stakeholders involved in the mobility ecosystem, from vehicle developers all the way to mobility-as-a-service companies and governmental authorities, to characterize rational trajectories for technology and investment development. We showcased our methodology on the real-world case study of Washington D.C., USA. Notably, we showed how our problem formulation allows for a systematic analysis of incomparable objectives, providing stakeholders with analytical insights for the socio-technical design of \glspl{abk:av}-enabled mobility systems.
This work opens the field for the following future research streams:\\
\emph{Modeling:} First, we would like to extend the presented framework to capture additional modes of transportation, such as buses, bikes, and e-scooters, and heterogeneous fleets with different self-driving infrastructures, propulsion systems, and passenger capacity. Second, we would like to include the possibility of accommodating the design of public transit lines. Third, we would like to investigate variable demand models. Finally, we would like to analyze the interactions between multiple stakeholders, characterizing the equilibrium arising from their conflicting interests.\\
\emph{Algorithms:} It is of interest to tailor general co-design algorithmic frameworks to the particular case of transportation design problems, possibly leveraging their specific structure.\\
\emph{Application:} Finally, we would like to devise a web interface which supports mobility stakeholders to reason on strategic interventions in urban areas.

\section{Acknowledgements}
We would like to thank Dr. Riccardo Bonalli, Dr. Maximilian Schiffer, Matthew Tsao, Dr. Kaidi Yang, and Dr. Stephen Zoepf for the fruitful discussions, Ms. Sonia Monti for the \gls{abk:iamod} illustration, and Dr. Ilse New for proofreading the manuscript. The first author would like to thank the Traugott and Josefine Niederberger-Kobold Foundation and FISITA for their financial support. The second author would like to thank the Zeno-Karl Schindler Foundation for the financial support.
This research was supported by the National Science Foundation under CAREER Award CMMI-1454737 and the Toyota Research Institute (TRI). This article solely reflects the opinions and conclusions of its authors and not NSF, TRI, or any other entity.

\section{Statement of Contributions}
The authors confirm contribution to the paper as follows: study conception and design: Gioele Zardini, Nicolas Lanzetti, Mauro Salazar, Andrea Censi, and Marco Pavone, data collection: Gioele Zardini, Nicolas Lanzetti, analysis and interpretation of results: Gioele Zardini, Nicolas Lanzetti, Andrea Censi, and Marco Pavone, draft manuscript preparation: Gioele Zardini, Nicolas Lanzetti, Mauro Salazar, Andrea Censi, Emilio Frazzoli, and Marco Pavone. 

\noindent All authors reviewed the results and approved the final version of the manuscript.

\newpage

\bibliographystyle{trb}
\bibliography{paper.bib}
\end{document}